\documentclass[useAMS,usenatbib]{mn2e}
\usepackage{graphicx}
\usepackage{latexsym}
\usepackage{psfig}
\usepackage{amssymb,amsmath}
\usepackage{subfig}
\usepackage{caption}
 \usepackage[T1]{fontenc}
 \usepackage{aecompl}

\title{Massive stars in massive clusters IV: Disruption of clouds by momentum--driven winds}

\author[J. E. Dale, J. Ngoumou, B. Ercolano, I.A. Bonnell]{J. E. Dale$^{1}$\thanks{E-mail: dale@usm.lmu.de (JED)}, J. Ngoumou$^{2}$, B. Ercolano$^{1}$, I. A. Bonnell$^{3}$\\
$^{1}$Excellence Cluster `Universe', Boltzmannstr. 2, 85748 Garching, Germany.\\
$^{2}$Universit\"{a}ts Sternwarte M\"{u}nchen, Scheinerstr. 1, 81679 M\"{u}nchen, Germany.\\
$^{3}$Department of Physics and Astronomy, University of St Andrews, North Haugh, St Andrews, Fife KY16 9SS}

\begin{document}

\pagerange{\pageref{firstpage}--\pageref{lastpage}} \pubyear{2006}

\maketitle

\label{firstpage}

\def\mnras{MNRAS}
\def\apj{ApJ}
\def\aj{AJ}
\def\aap{A\&A}
\def\apjl{ApJL}
\def\apjs{ApJS}
\def\araa{ARA\&A}
\def\pasp{PASP}
 
\begin{abstract}
We examine the effect of momentum--driven OB--star stellar winds on a parameter space of simulated turbulent Giant Molecular Clouds using SPH hydrodynamical simulations. By comparison with identical simulations in which ionizing radiation was included instead of winds, we show that momentum--driven winds are considerably less effective in disrupting their host clouds than are HII regions. The wind bubbles produced are smaller and generally smoother than the corresponding ionization--driven bubbles. Winds are roughly as effective in destroying the very dense gas in which the O--stars are embedded, and thus shutting down the main regions of star--forming activity in the model clouds. However, their influence falls off rapidly with distance from the sources, so they are not as good at sweeping up dense gas and triggering star formation further out in the clouds. As a result, their effect on the star formation rate and efficiency is generally more negative than that of ionization, if they exert any effect at all.\\
\end{abstract}

 
\section{Introduction}
The dispersal of molecular clouds and embedded clusters, and the triggering and suppression of star formation are still the subject of lively discussion and debate. Most clusters shed the remains of their natal gas clouds very early on, resulting in star formation efficiencies of the order of one percent \citep[e.g][]{2003ARA&A..41...57L}. Several internal feedback mechanisms arising from stars have been invoked as means of destroying GMCs while they are still forming stars. On the scale of molecular clouds, the most important are likely to be HII regions, stellar winds and supernovae from OB--type stars, but the relative contributions of these mechanisms is still debated.\\
\indent Early work by \cite{1984ApJ...278L.115M} suggested that wind bubbles were likely to be confined by HII regions and more recent numerical simulations by \cite{2003ApJ...594..888F,2006ApJ...638..262F} confirm that, except for very massive stars, wind bubble expansion driven by single stars is likely to stall inside the HII region where the pressure in the wind bubble becomes equal to the pressure in the ionized gas. The analysis of \cite{2002ApJ...566..302M} also concluded that expanding HII regions are much more important feedback agents than winds or supernovae.\\
\indent From the numerical perspective, there have been numerous studies done on GMC--scales of the interaction of clouds with internally--driven HII regions. Much early work concentrated on the modelling of ionization--driven champagne flows \citep[e.g.][]{1979A&A....71...59T,1979ApJ...233...85B,1979MNRAS.186...59W,1997ApJ...476..166W} and showed that photoevaporation could be an effective dispersal mechanism of uniform clouds on $10^{7}$yr timescales if the massive stars were located near the peripheries of the clouds. However, \cite{1980A&A....90...65M} and \cite{1989A&A...216..207Y} showed that the dispersal efficiency was reduced by the action of gravity or by placing the stars deep inside the clouds. \cite{1997ApJ...476..166W} and \cite{2002ApJ...566..302M} extended the work of \cite{1979MNRAS.186...59W}. They noted that the ionizing feedback from a single OB--association typically operates for 3--4 Myr and that a single such association will not be able to destroy a GMC. \cite{1997ApJ...476..166W} found that high--mass clouds ($M\gtrsim3\times10^{5}$M$_{\odot}$) could be photoevaporated by the combined action of many blister HII regions driven by OB associations over $\sim$30 Myr, but that small clouds were more likely to be disrupted, rather than ionized. \cite{2002ApJ...566..302M} obtained similar results, although they derived somewhat shorter destruction timescales from blister HII regions of 17--24 Myr for 10$^{6}$M$_{\odot}$ clouds.\\
\indent More recent work has examined the influence, in more complex environments, of HII regions \citep[e.g][]{2005MNRAS.358..291D,2010ApJ...719..831P,2010ApJ...715.1302V,2011MNRAS.414..321D,2012MNRAS.424..377D}, jets \citep[e.g][]{2006ApJ...640L.187L,2010ApJ...709...27W,2012ApJ...747...22H,2012ApJ...754...71K}, and winds \citep{2008MNRAS.391....2D,2010MNRAS.405..401D,2012ASPC..453...25F,2013arXiv1302.2443R}. In general, these feedback mechanisms seem to be able to reduce star formation efficiencies or rates, but not by very large factors, and, as the earlier studies of champagne flows found, success in disrupting or dispersing clouds depends rather strongly on the cloud properties. In particular, \cite{2012ASPC..453...25F} showed explicitly that wind--driven champagne flows are much more destructive to uniform clouds than to structured clouds. Structured clouds very often provide avenues of escape for $>10^{6}$K wind gas, allowing it to exit the cloud and decreasing its influence on the cold cloud material.\\
\indent Most of these simulations concentrate on only one form of feedback. It is prudent to introduce new physical effects to simulations one at a time and to explore how each behaves on its own before attempting calculations in which all effects are included, as these are likely to be very complicated and difficult to interpret. In this paper, we concentrate on the effects of main--sequence winds alone from OB stars on molecular clouds. We have three objectives: Firstly, we want to understand the reaction of turbulent clouds to momentum input from their stars. Secondly, we wish to perform a controlled experiment to test the contention that the dynamical influence of winds on GMCs is considerably smaller than that of photoionization. We will achieve this by direct comparison of the simulations presented here with our recent work in which we examined the impact of photoionizing radiation from OB--stars on a parameter--space of model GMCs \citep[][hereafter Papers I and II]{2012MNRAS.424..377D,2012arXiv1212.2011D}. Thirdly, these simulations will provide a baseline for later work in which we will include winds \emph{and} ionization, so that we can isolate the effects of the two different processes.\\
\indent In the next section, we briefly discuss the physics of stellar winds. Our numerical setup is described in Section 3. Section 4 contains our results, and our discussion and conclusions follow in sections 5 and 6.\\
\section{Stellar winds}
\indent Winds inject mass, momentum and energy into GMCs. The mass they inject is dynamically unimportant, but the momentum and energy carried by the injected mass can have profound effects on the host clouds. While the momentum emitted by the massive stars is necessarily absorbed by the clouds (except along lines of sight on which the winds intercept no cloud material), the fate of the energy is much more difficult to determine.\\
\indent We stress here that by `winds', we mean line--driven spherically--symmetric (or nearly so) outflows originating from close to the surface of a given massive star once it has reached the main sequence, and suffering no intrinsic collimation by, e.g., circumstellar disks. We do not include the kind of feedback variously referred to as protostellar winds/jets/outflows, which are associated with disk accretion and are usually collimated to some degree by the circumstellar magnetic field. Feedback of this nature also injects large quantities of momentum into clouds, over shorter timescales than main--sequence winds, but at higher rates. Several authors have recently concluded that protostellar outflows are able to drive or maintain turbulence on $\sim$pc scales in embedded clusters \cite{2006ApJ...640L.187L,2007ApJ...662..395N,2007ApJ...659.1394M,2009ApJ...695.1376C,2010ApJ...709...27W} (although \cite{2007ApJ...668.1028B} reached a different conclusion). It is also likely that the action of collimated outflows would modify the effects of other modes of feedback by, for example, punching low optical depth channels through the circumstellar material \citep[e.g.][]{2011ApJ...740..107C}. We neglect collimated outflows in this work because, while they clearly affect the dynamics at size scales of pc, mass scales of $\lesssim10^{3}$M$_{\odot}$ \citep{2007ApJ...659.1394M} and timescales of $\lesssim10^{6}$ yr, their influence on clouds of the masses and sizes considered here over the timescales for which our simulations run are likely to be small compared to other forms of feedback, in particular HII regions and main--sequence winds. From a more technical point of view, it would be difficult for us to include such feedback in our models anyway since, in many of our simulations, we do not resolve individual stars, so that the direction and strength of the outflows would have to be set rather arbitrarily. Additionally, since collimated outflows from all stellar masses, and not just the massive stars, would probably need to be included, the simulations would likely be prohibitively expensive to run over the $\sim$3Myr pre--supernova timespan of interest here. The effect of collimated outflows on other forms of feedback does, however, need to be examined, but is outside the remit of this paper.\\
\indent The interaction of spherically--symmetric main--sequence O--star winds with clouds was first studied in detail by \cite{1977ApJ...218..377W}. The kinetic energy contained in the wind is rapidly thermalized when the wind collides with cooler, denser material at the edge of the wind bubble (either the inner face of an HII region, or sometimes cold molecular gas). The very hot ($>10^{6}$K) shocked wind may then cool by thermal conduction or mixing across the contact--discontinuity (cooling can also occur in the bubble interior if surviving clumps are able to mass--load the wind, \cite{2005MNRAS.361.1077P}). The behaviour of the wind bubble then depends on the cooling timescale set by these process compared with the expansion timescale of the bubble (see \cite{2001PASP..113..677C} for a detailed discussion of this issue). The extremal assumptions that can be made are that the cooling is slow, so that the wind bubble behaves adiabatically and its expansion is driven by the thermal pressure in the shocked wind (and its radius evolves with time as $R(t)\propto t^{3/5}$ in a uniform medium), or that the cooling is very fast, so that the bubble behaves isothermally and is driven by simple ram pressure ($R(t)\propto t^{1/2}$ in a uniform medium).\\
\indent \cite{1992ApJ...388...93K} and \cite{1992ApJ...388..103K} discuss in detail the intermediate case of partially radiative bubbles (PRBs), where most of the wind has cooled and collapsed into a thin shell lining the shell of swept--up ambient medium, but the most recently emitted wind has not yet cooled and fills most of the bubble volume. For a main--sequence O--star wind blowing in a uniform medium, this phase is never achieved before the bubble comes into pressure equilibrium with the external medium, so the bubble always behaves adiabatically in this case, unless there is additional mass injected inside the bubble by, for example, evaporating clumps. However, in the case of medium whose density falls off as $r^{-2}$, which is the case in our simulated clouds, the expansion of an adiabatic bubble should accelerate. This would make the bubble Rayleigh--Taylor unstable, which would likely lead to mixing at the interface between the cool swept--up material and the wind, resulting in cooling of the wind.\\
\indent It might be supposed that the issue of the temperatures inside wind bubbles could be easily solved observationally, since the temperatures generated by thermalizing the kinetic energy transported by 1 000 km s$^{-1}$ flows should result in X--ray--emitting diffuse gas. However, observing such emission is very difficult because the young star--forming regions in which it would be found are teeming with low--mass YSOs. These objects are also X--ray sources and extracting them from images is very difficult \citep[e.g][]{2006AJ....131.2140T}. Even when diffuse X--ray emission can be detected, it is also not always possible to establish whether its origin is winds or supernovae.\\
\indent Diffuse X--ray emission likely produced by wind thermalization is detected in M17 and in the Rosette Nebula \citep[e.g][]{2003ApJ...593..874T}. However, \cite{2009ApJ...693.1696H} compared the \cite{1977ApJ...218..377W} model to the Carina nebula and concluded that the observed X--ray luminosity is two orders of magnitude lower than the model predicts and that the high filling factor of photoionized gas implies that the dynamics of the bubbles are controlled by HII regions and not winds.\\
\indent Given the theoretical and observational uncertainties, we choose to model winds in a very simple way, by treating the O--stars as sources of momentum only, so that we are effectively modelling a lower limit to how much damage winds can do to GMCs by ignoring wind thermal pressure and ablation. As we show later, this assumption is likely to be reasonable in the highly inhomogeneous ambient environments of interest here. Recent work by \cite{2012ASPC..453...25F}and \cite{2013arXiv1302.2443R} using Eulerian numerical schemes, found that the very hot wind gas rapidly escapes from clumpy molecular clouds and that the cold gas is able to resist ablation for several Myr.\\
\begin{figure}
\includegraphics[width=0.5\textwidth]{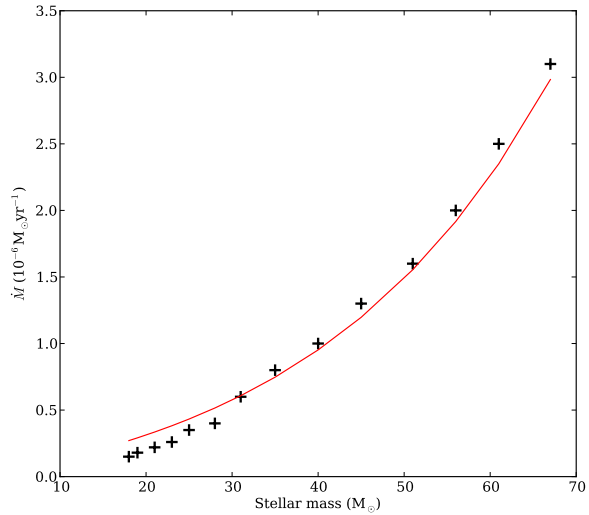}
\caption{Plot of $\dot{M}$ in units of 10$^{-6}$M$_{\odot}$yr$^{-1}$ from Smith et al (2006) (black crosses) with the fit given in Equation \ref{eqn:mdot} shown as a red line.} 
\label{fig:mdotfit}
\end{figure}
\begin{figure}
\includegraphics[width=0.5\textwidth]{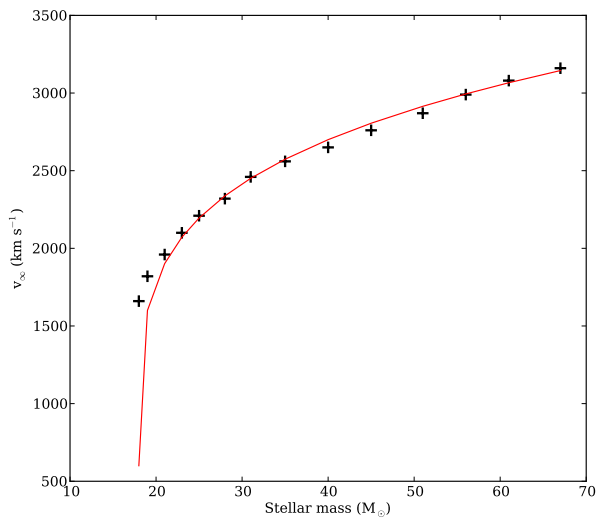}
\caption{Plot of v$_{\infty}$ in units of km s$^{-1}$ from Smith et al (2006) (black crosses) with the fit given in Equation \ref{eqn:vinf} shown as a red line.} 
\label{fig:vinffit}
\end{figure}
\section{Numerical methods}
\indent We use very similar numerical techniques and identical initial conditions to those presented in our earlier studies of the effects of photoionizing feedback on a parameter space of turbulent molecular clouds \citep{2012MNRAS.424..377D,2012arXiv1212.2011D}. We use a well--known variant of the Benz \citep{1990nmns.work..269B} Smoothed Particle Hydrodynamics \citep{1992ARA&A..30..543M} code, which is ideal for studying the evolution of molecular clouds and embedded clusters. In all our simulations, we begin with 10$^{6}$ gas particles.  We use the standard artificial viscosity prescription, with $\alpha=1$, $\beta=2$. Particles are evolved on individual timesteps. The code is a hybrid N--body SPH code in which star formation is modelled using point--mass sink particles \citep{1995MNRAS.277..362B}. Self--gravitational forces between gas particles are calculated using a binary tree, whereas gravitational forces involving sink--particles are computed by direct summation. Sink particles are formed dynamically and may accrete gas particles and grow in mass. In our simulations of 10$^{5}$ and 10$^{6}$M$_{\odot}$ clouds, the sink particles represent stellar clusters, since the mass resolution is not sufficient to capture individual stars. Clusters approaching each other to within their accretion radii are merged if they are mutually gravitationally bound. In our 10$^{4}$M$_{\odot}$ simulations, sink particles represent individual stars. Their accretion radii are set to 0.005pc ($\sim10^{3}$ AU) and mergers are not permitted. In all simulations gravitational interactions of sink particles with other sink particles are smoothed within their accretion radii.\\
\indent For consistency with our earlier work, we treat the thermodynamics of the neutral gas using a piecewise barotropic equation of state from \cite{2005MNRAS.359..211L}, defined so that $P = k \rho^{\gamma}$, where
\begin{eqnarray}
\begin{array}{rlrl}
\gamma  &=  0.75  ; & \hfill &\rho \le \rho_1 \\
\gamma  &=  1.0  ; & \rho_1 \le & \rho  \le \rho_2 \\
\gamma  &=  1.4  ; & \hfill \rho_2 \le &\rho \le \rho_3 \\
\gamma  &=  1.0  ; & \hfill &\rho \ge \rho_3, \\
\end{array}
\label{eqn:eos}
\end{eqnarray}
and $\rho_1= 5.5 \times 10^{-19} {\rm g\ cm}^{-3} , \rho_2=5.5 \times10^{-15} {\rm g cm}^{-3} , \rho_3=2 \times 10^{-13} {\rm g\ cm}^{-3}$. At low densities, $\gamma$ is less than unity, implicitly ensuring that the gas at low densities is warmer than the canonical temperature of $\sim10$K. The isothermal $\gamma=1.0$ segment at moderate densities approximates the effect of dust cooling and the $\gamma=1.4$ segment represents the regime where dense collapsing cores become optically thick and behave adiabatically. The final isothermal phase of the equation of state is simply in order to allow sink-particle formation to occur. Once the minimum gas temperature, which we set to 7.5K, is specified, the relation between $\rho$ and $T$ is fixed. All our simulated clouds have initial average densities $< \rho_{1}$, so that they lie in the line--cooling regime with temperatures in excess of 10K.\\
\indent Our model clouds initially have a Gaussian three--dimensional density profile with the ratio of the maximum and minimum gas densities being approximately three. This is done simply to ensure that the clouds remain centrally condensed in the early stages of their evolution while the dynamics is dominated by turbulence. We seed the gas with a Kolmogorov turbulent velocity field such that the clouds have initial virial ratios close to 0.7 (i.e. bound clouds) or 2.3 (unbound clouds). The clouds are allowed to evolve and convert gas into stellar material until a few stars have masses in excess of 20$M_{\odot}$ (for clouds with masses $\leq3\times10^{4}$M$_{\odot}$ where individual stars can be resolved) or a few clusters sufficiently massive to host an O--star (in simulations of more massive clouds). At this point, which different clouds take very different times to reach (from $<2$Myr for Run UP to $\approx20$Myr for Run A), winds are enabled for these sources and any other sink particles that grow to be sufficiently massive later. The simulations are then continued for as near as possible to 3Myr. The intent is to model the action of winds from the time when the first few massive stars form to the time when the first of them is likely to explode as a supernova. For convenience, Table \ref{tab:runs} gives the most important parameters of all simulations from Papers I and III for which we have simulated the effects of winds.\\
\begin{table*}
\begin{tabular}{|l|l|l|l|l|l||l|l|l|}
Run&Mass (M$_{\odot}$)&Radius (pc)&v$_{\rm RMS}$ (km s$^{-1}$)&v$_{\rm ESC}$ (km s$^{-1}$)&$\langle$ n(H$_{2}$) $\rangle$ (cm$^{-3}$)&t$_{\rm ff}$ (Myr)&E$_{\rm kin}$/$|$E$_{\rm pot}|$\\
\hline
A&$10^{6}$&180&5.0&6.9&2.9&19.6&0.7\\
\hline
B&$10^{6}$&95&6.9&9.5&16&7.50&0.7\\
\hline
D&$10^{5}$&45&3.0&4.4&15&7.70&0.7\\
\hline
E&$10^{5}$&21&4.6&6.4&147&2.46&0.7\\
\hline
I&$10^{4}$&10&2.1&2.9&136&2.56&0.7\\
\hline
J&$10^{4}$&5&3.0&4.1&1135&0.90&0.7\\
\hline
\hline
UB&$3\times10^{5}$&45&10.0&7.6&45&6.0&2.3\\
\hline
UV&$10^{5}$&21&8.4&6.4&148&3.3&2.3\\
\hline
UU&$10^{5}$&10&12.3&9.3&1371&1.1&2.3\\
\hline
UF&$3\times10^{4}$&10&6.7&5.1&410&2.0&2.3\\
\hline
UP&$10^{4}$&2.5&7.6&5.9&9096&0.4&2.3\\
\hline
UQ&$10^{4}$&5.0&5.4&4.1&1137&1.2&2.3\\
\end{tabular}
\caption{Initial properties (mass, radius, initial turbulent velocity dispersion, escape velocity, mean initial molecular number density, freefall time, and initial virial ratio) of the runs from Papers I and II for which we have simulated the effects of winds.}
\label{tab:runs}
\end{table*}
\indent We simulate the action of stellar winds by injecting momentum isotropically from sink particles in the manner described in \cite{2008MNRAS.391....2D}. The winds algorithm presented and tested in that work is unmodified here save in one respect: instead of using Equation 2 from \cite{2008MNRAS.391....2D} to determine wind mass loss rates and adopting a fixed value of $v_{\infty}$ to obtain momentum fluxes, we instead use approximate fits to the data supplied by \cite{2006ApJ...644.1151S} on stars in the Carina Nebula. We convert spectral types to masses using the table presented in \cite{2010A&A...524A..98W} and fit the mass loss rate $\dot{M}(M_{*})$ and $v_{\infty}(M_{*})$ by the analytic formulae
\begin{eqnarray}
\dot{M}(M_{*})=\left[0.3{\rm ~exp}\left(\frac{M_{*}}{28}\right)-0.3\right]\times10^{-6}{\rm M}_{\odot}{\rm yr}^{-1}
\label{eqn:mdot}
\end{eqnarray}
and
\begin{eqnarray}
v_{\infty}(M_{*})=\left[10^{3}(M_{*}-18)^{0.24}+600\right]{\rm km~s}^{-1}
\label{eqn:vinf}
\end{eqnarray}
In Figures \ref{fig:mdotfit} and \ref{fig:vinffit} we compare these fits to the data from \cite{2006ApJ...644.1151S}.\\
\indent Note that we neglect stellar evolution in computing wind mass--loss rates and velocities (as we did in our previous work when computing ionizing luminosities). During the RSG, WR and LBV phases, the wind momentum fluxes may change substantially. Particularly in the WR phase, the momentum flux can be considerably larger than when the star is on the main sequence, but the WR phase is also much shorter than the MS phase, so the total quantity of momentum injected in the two phases is likely to be comparable, and including these later phases of evolution will not greatly alter the momentum--budget of our model clouds. In any case, it is likely that the influence of winds will be dominated by their early, i.e. main--sequence--driven, phases. For a given sink particle, the mass--loss rate and wind terminal velocity are assumed to be functions of mass only and change only if the sink accretes more material.\\
\indent We use the same criteria as in our earlier work to determine which sink particles be treated as feedback sources and the times when winds are activated in these calculations are exactly the same as the times when ionization is enabled in Papers I and II, allowing detailed quantitative comparisons of the two different forms of feedback to be made.\\
\section{Results}
\subsection{Changes in the gas structure}
\begin{figure*}
  \captionsetup[subfigure]{labelformat=empty}
     \centering
     \subfloat[]{\includegraphics[width=0.95\textwidth]{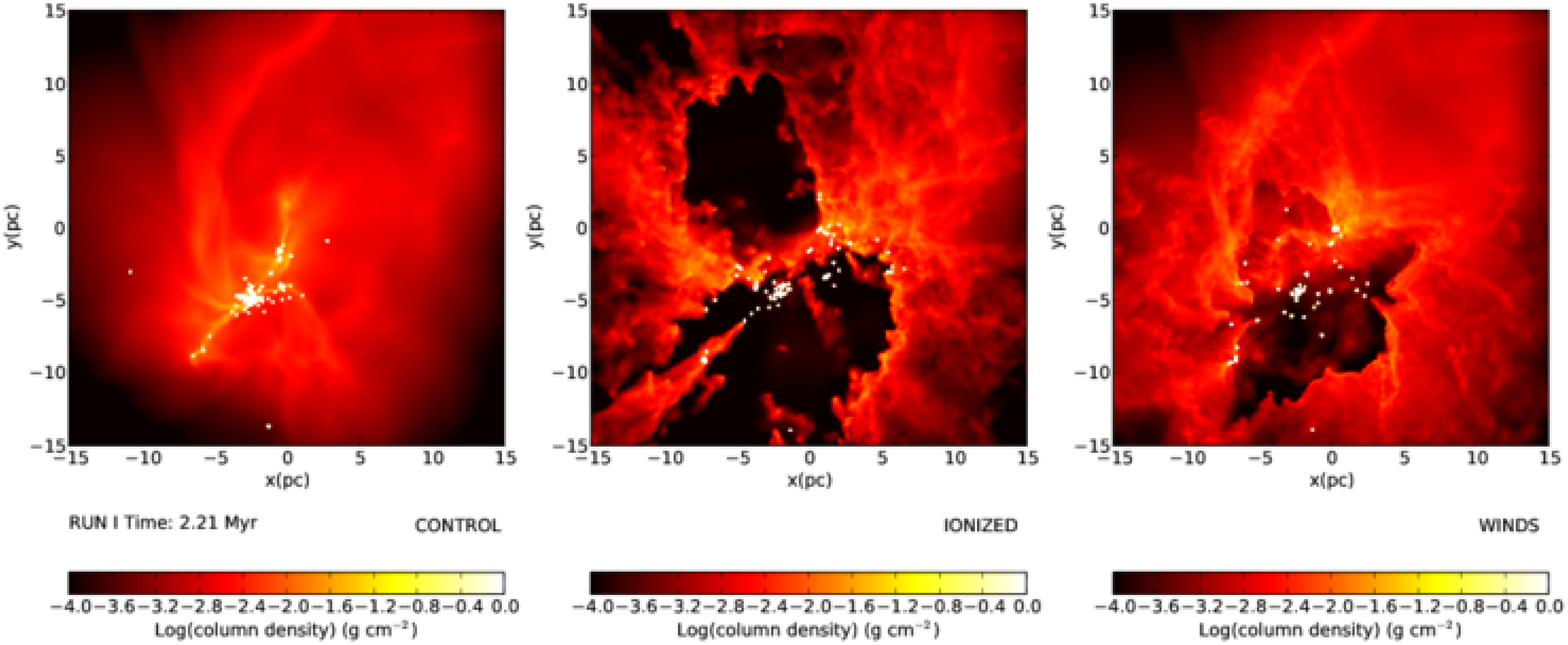}}     
     \vspace{-0.25in}
     \subfloat[]{\includegraphics[width=0.95\textwidth]{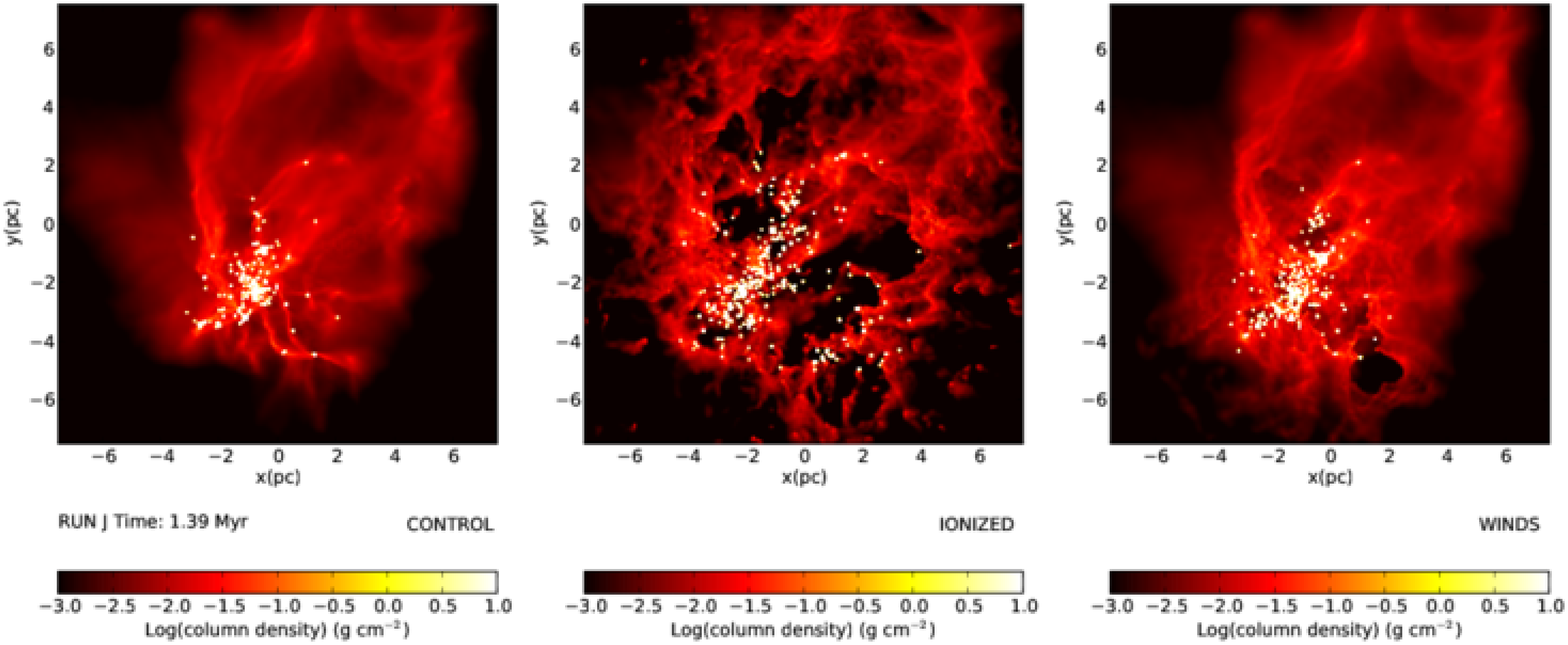}}
     \vspace{-0.25in}
      \subfloat[]{\includegraphics[width=0.95\textwidth]{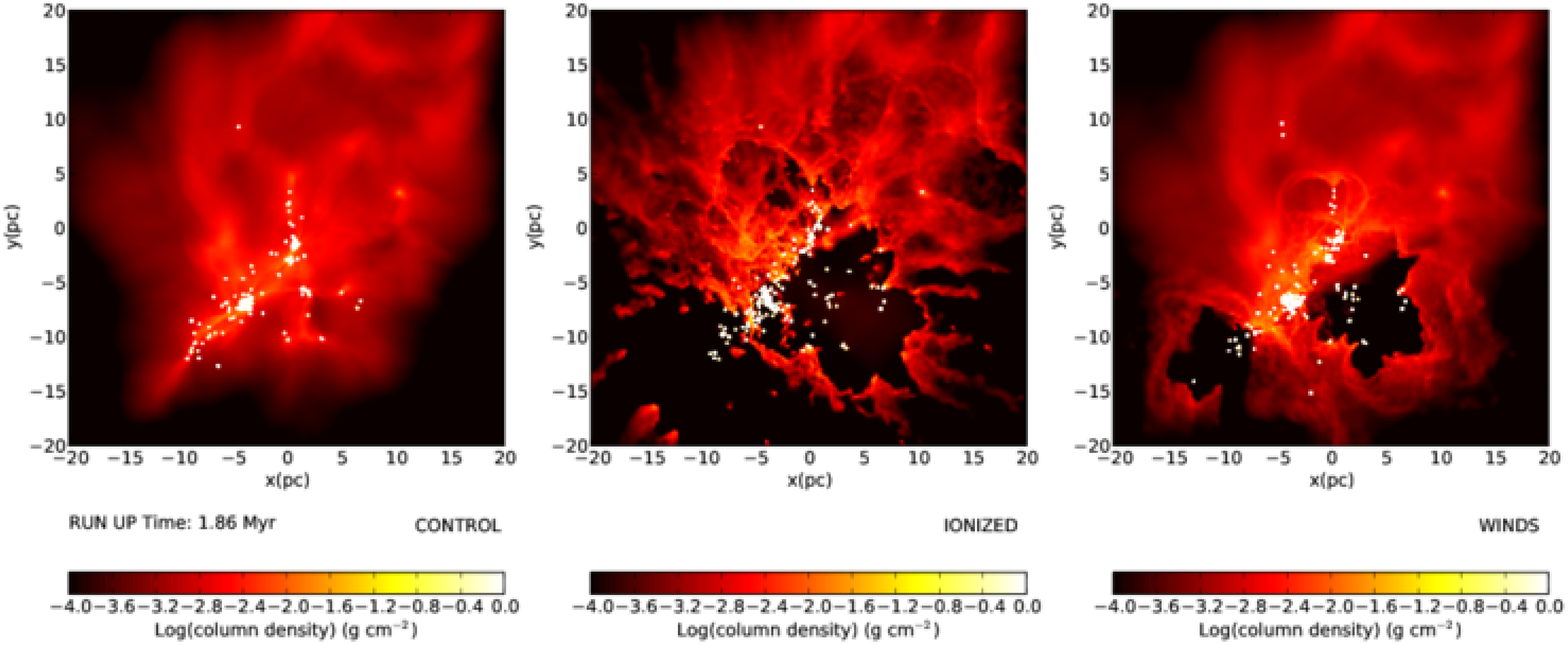}}
\caption{Column density plot viewed along the z--axis of the control (left panel), ionized (centre panel) and windblown (right panel) simulations I, J and UP.} 
\label{fig:compare_bnd}
\end{figure*}

\begin{figure*}
  \captionsetup[subfigure]{labelformat=empty}
     \centering
      \subfloat[]{\includegraphics[width=0.95\textwidth]{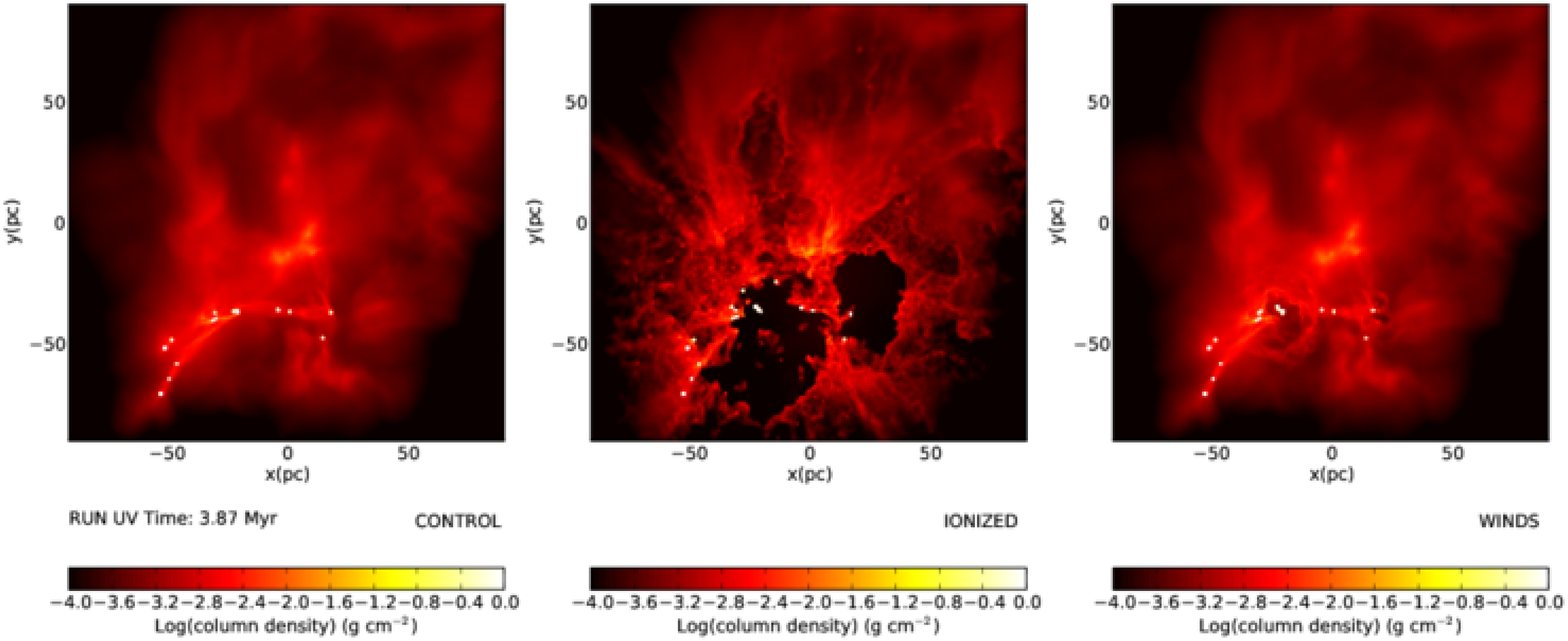}}
     \vspace{-0.25in}
     \subfloat[]{\includegraphics[width=0.95\textwidth]{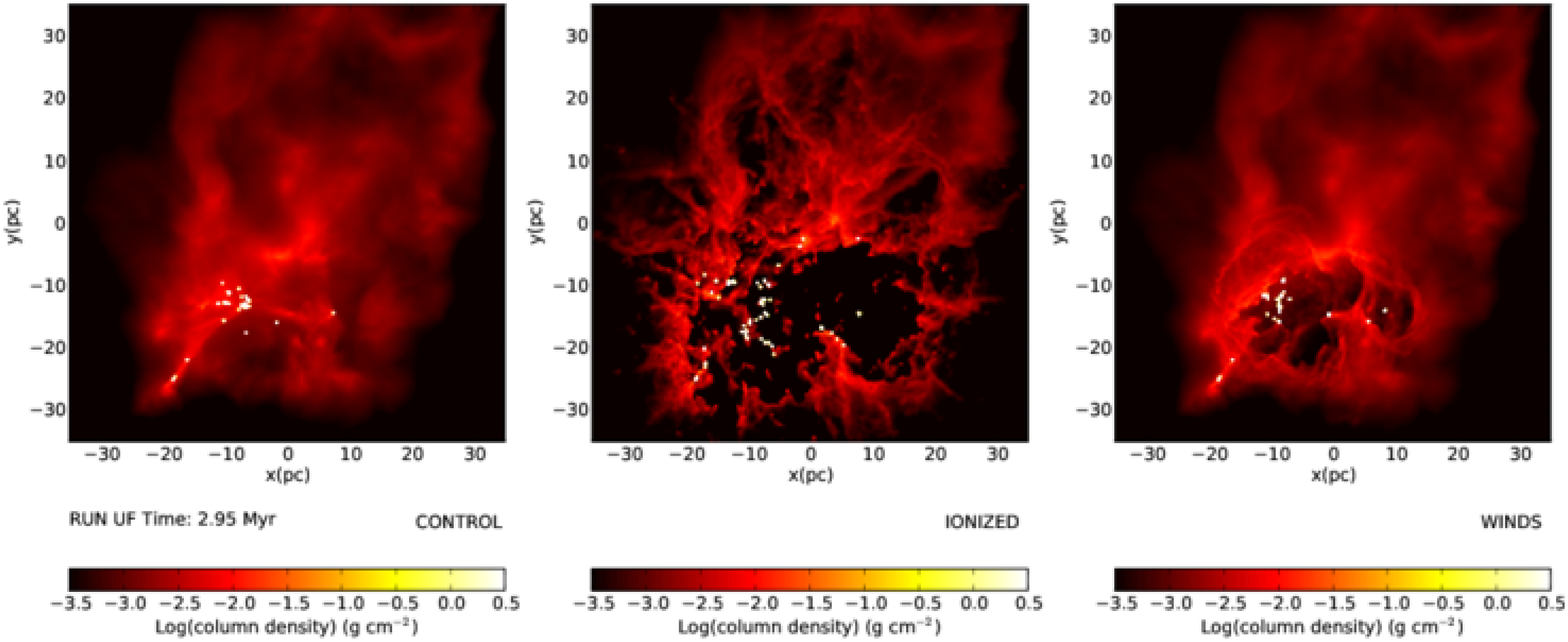}}
     \vspace{-0.25in}     
     \subfloat[]{\includegraphics[width=0.95\textwidth]{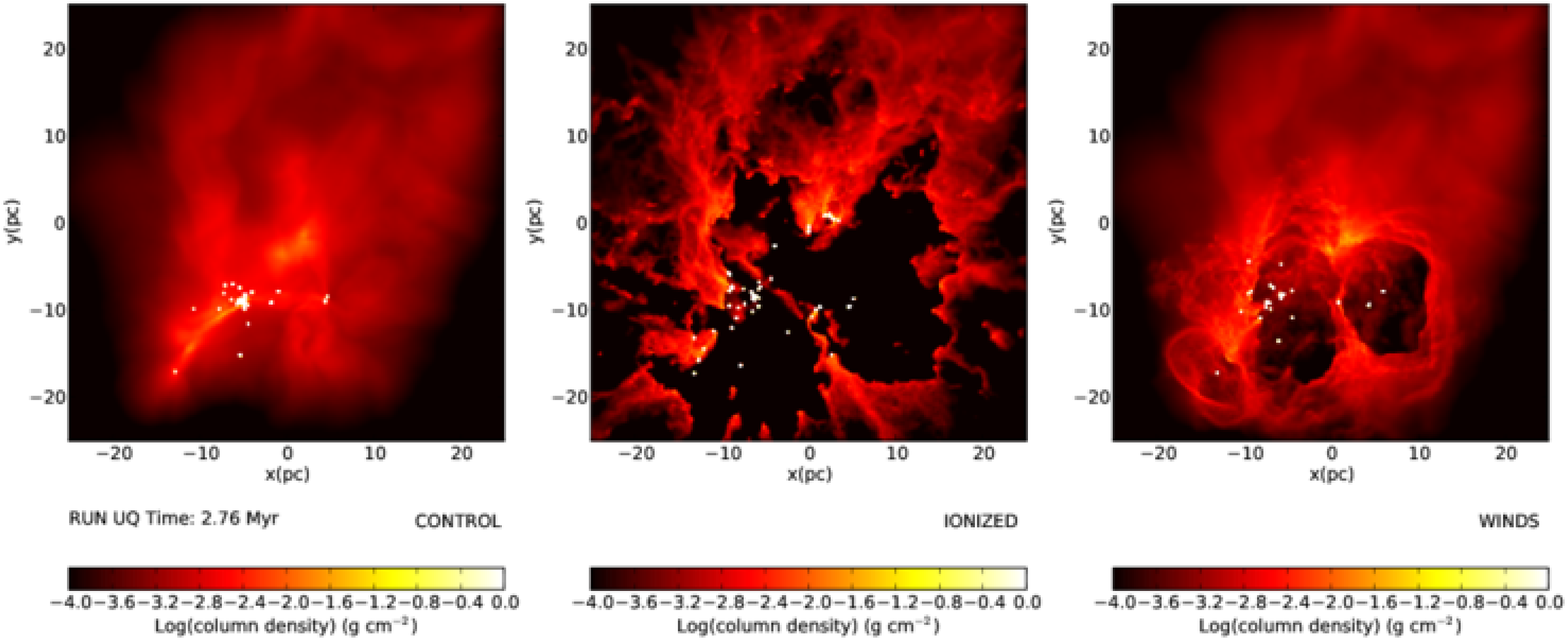}}
\caption{Column density plot viewed along the z--axis of the control (left panel), ionized (centre panel) and windblown (right panel) unbound--cloud simulations UV, UF and UQ.} 
\label{fig:compare_unbnd}
\end{figure*}

In Figures \ref{fig:compare_bnd} and \ref{fig:compare_unbnd}, we show column density plots as viewed along the z--axis of a selection of simulations in which there is either no feedback at all (left panels), ionizing feedback \emph{only} (centre panels) or wind feedback \emph{only} (right panels). We selected the six simulations in which winds have the strongest influence on the dynamics of the clouds. We also simulated the effects of winds on most of the other clouds from Papers I and II, including Runs A, B, D, E, UB, and UU, but in these calculations, the dynamical effects of winds on the cloud evolution was essentially negligible, particularly for the high mass clouds. This result is discussed in detail in Section 5.\\
\indent Figures \ref{fig:compare_bnd} and  \ref{fig:compare_unbnd} clearly show that the effects of photoionization and winds differ in both manner and degree. The windblown bubbles are smoother and tend to be rounder than those generated by ionization, and there are also several smooth shell--like structures present in the wind simulations which do not occur in the ionized runs. In addition, the ionized bubbles plainly occupy much larger volumes of the clouds and many have in fact broken out of the clouds entirely, which is less often the case for the wind bubbles. It therefore appears that, in a given time period, winds are able to do considerably less damage to the clouds than ionization can.\\
\subsection{Relative ability of HII regions and winds to disrupt clouds}
\begin{figure*}
     \centering
     \subfloat[Run I]{\includegraphics[width=0.30\textwidth]{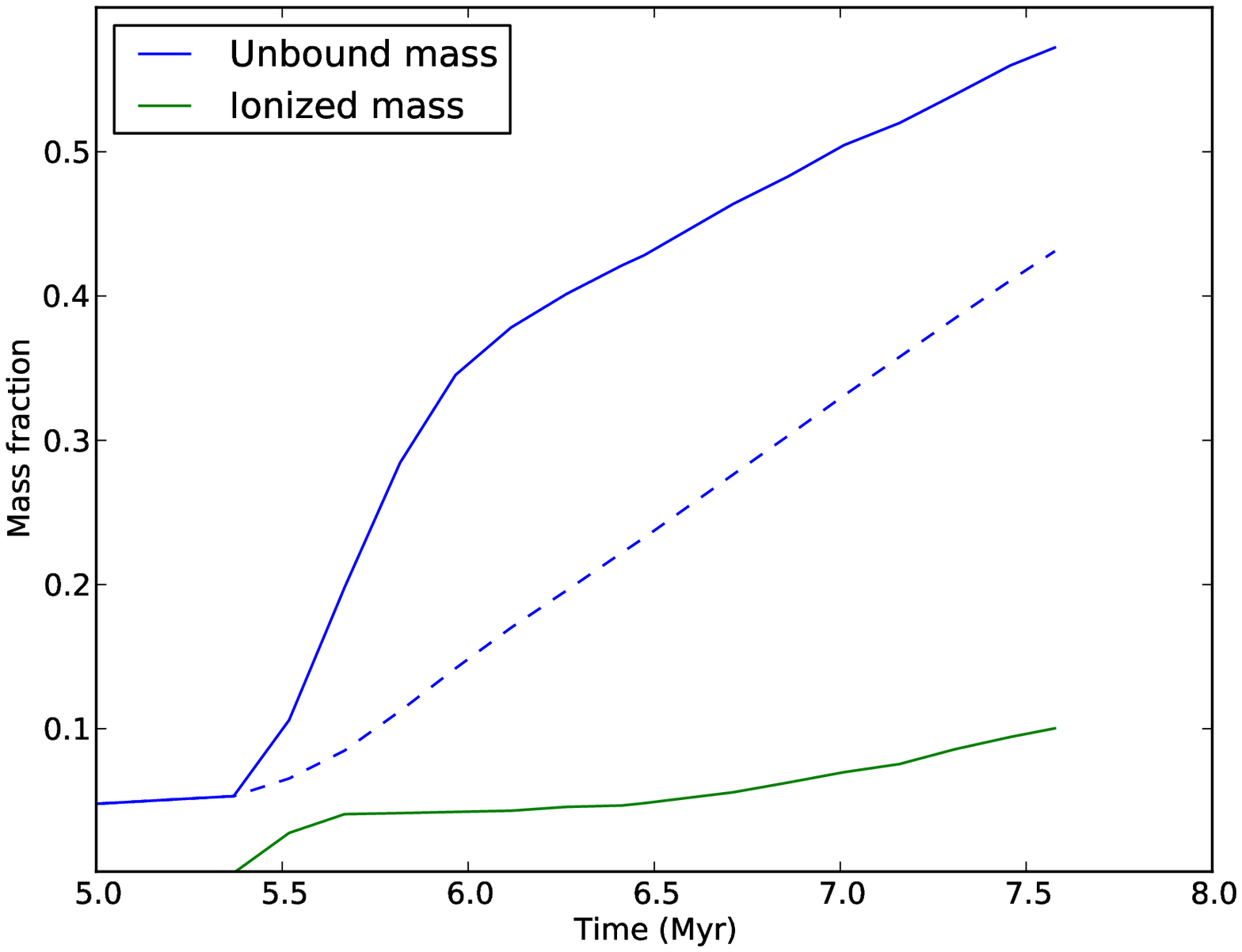}}
     \hspace{.01in}
    \subfloat[Run J]{\includegraphics[width=0.30\textwidth]{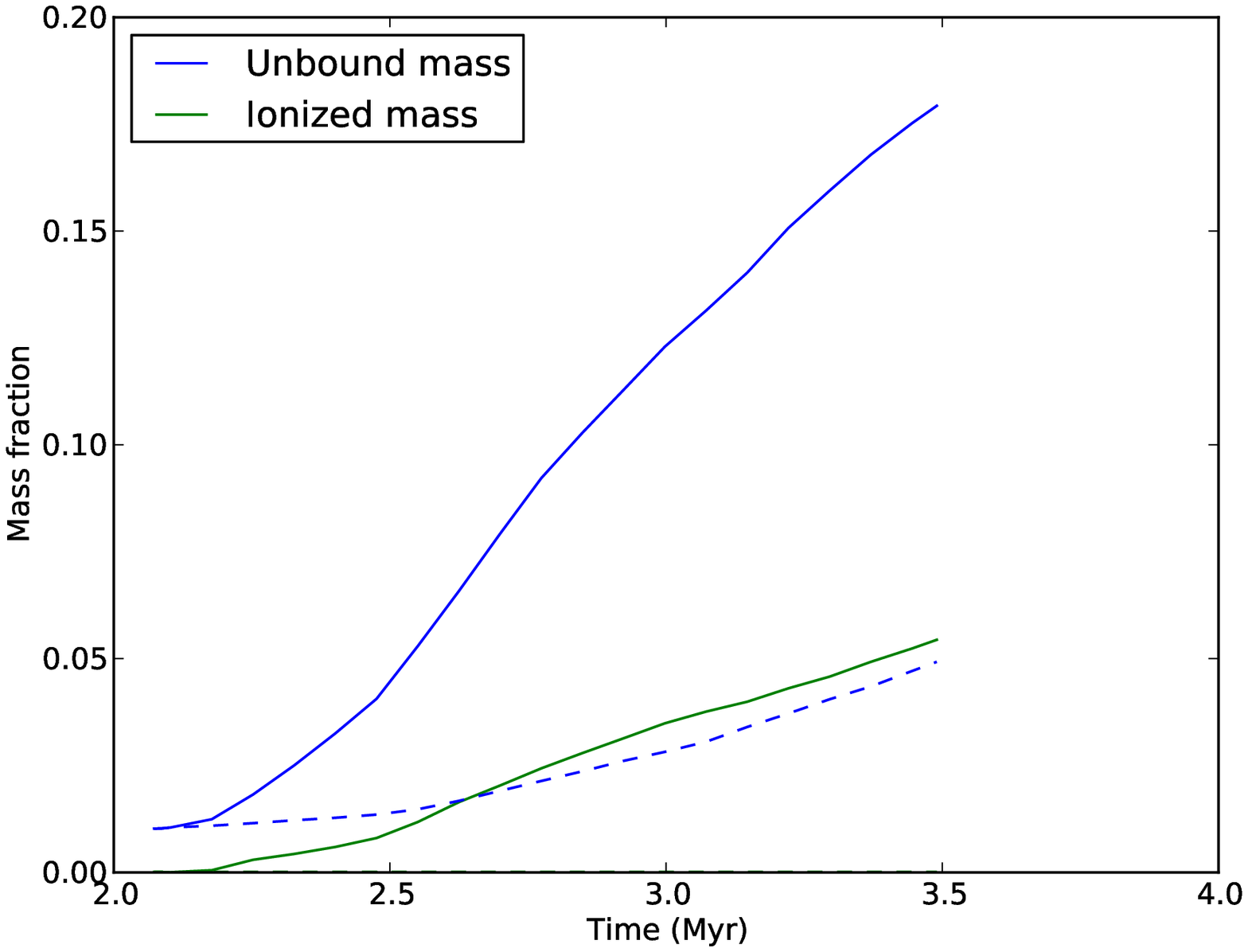}}
    \hspace{.01in}
    \subfloat[Run UP]{\includegraphics[width=0.30\textwidth]{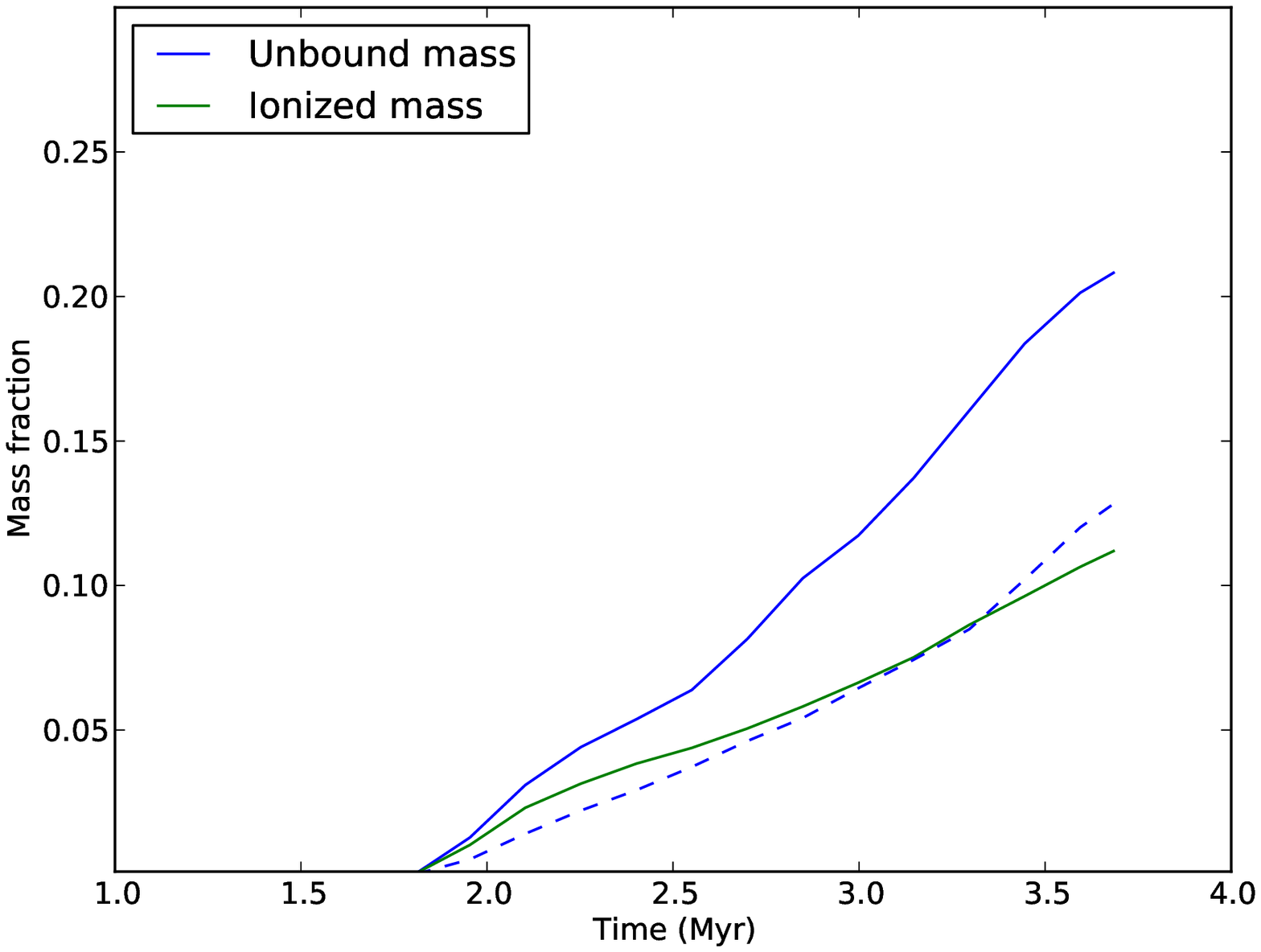}}
    \vspace{.01in}
     \subfloat[Run UV]{\includegraphics[width=0.30\textwidth]{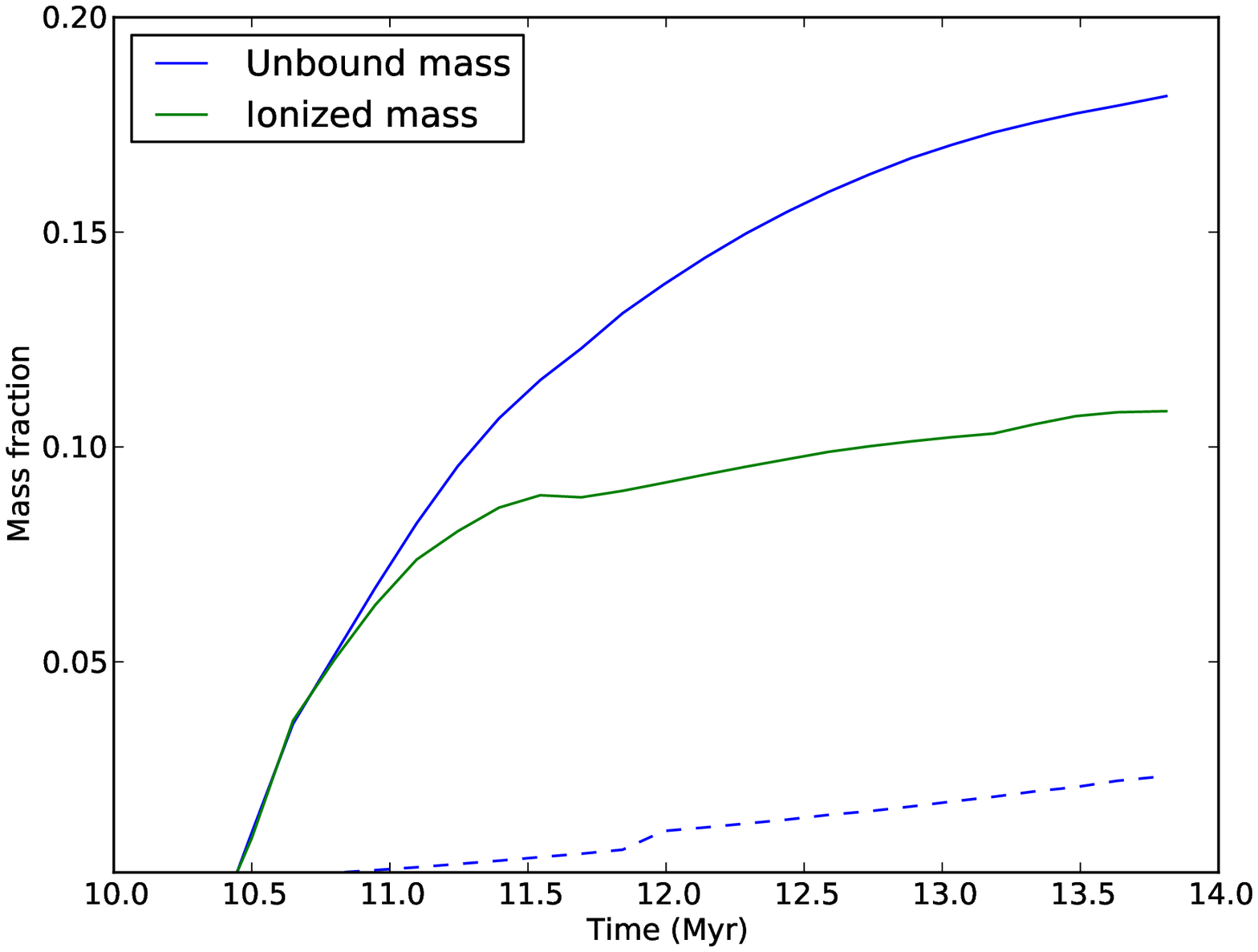}}     
     \hspace{.01in}
     \subfloat[Run UQ]{\includegraphics[width=0.30\textwidth]{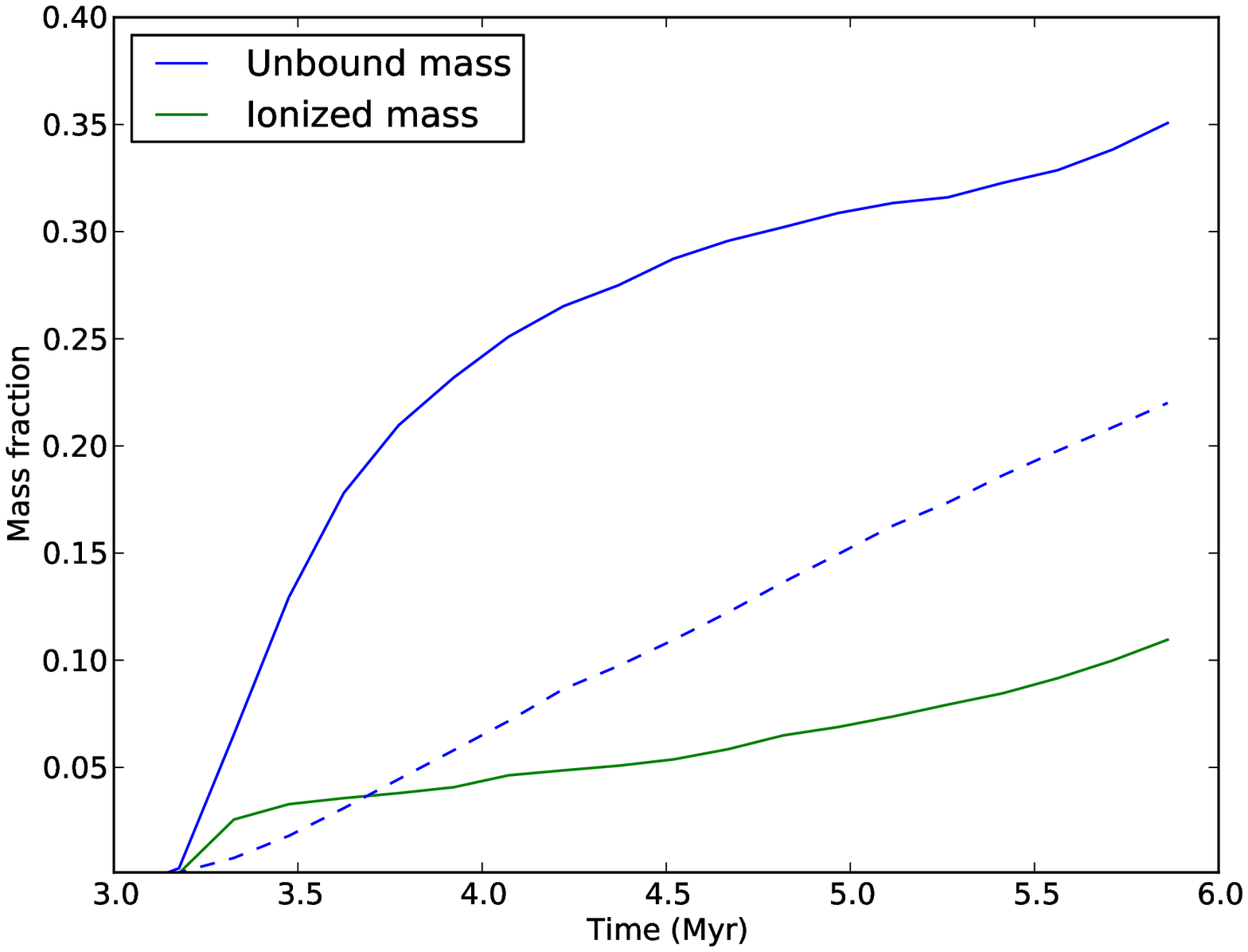}}
     \vspace{.01in}
     \subfloat[Run UF]{\includegraphics[width=0.30\textwidth]{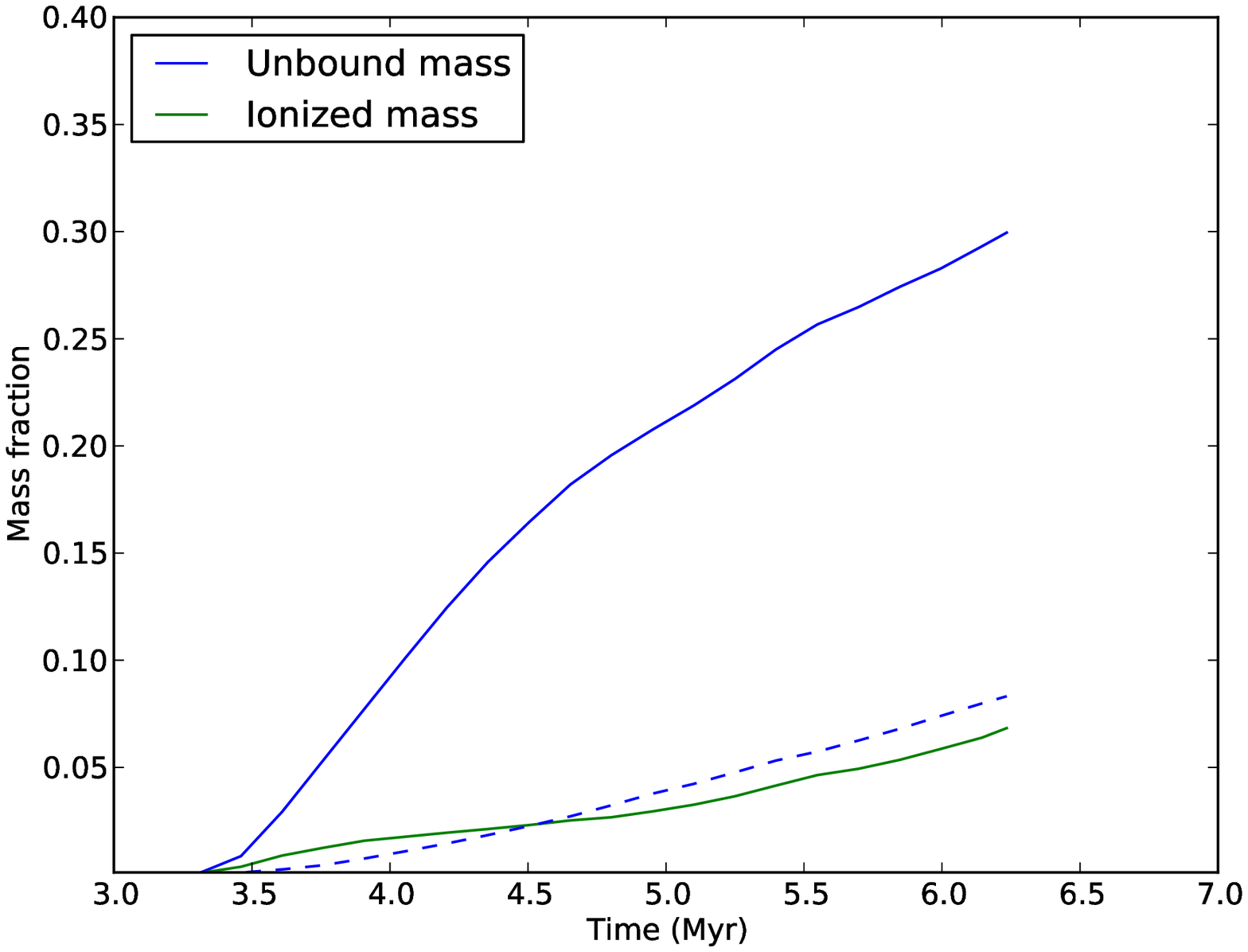}}
\caption{Time variation of the unbound mass fractions in the ionized (blue solid lines) and wind (blue dashed lines) and of the ionization fractions from the ionized simulations (green solid lines) in simulations I, J and UP (top row), and UV, UQ and UF (bottom row).} 
\label{fig:disrupt}
\end{figure*}

Figure \ref{fig:disrupt} depicts the evolution with time of the unbound gas fraction (blue lines) from the six selected  windblown simulations (dashed lines) and the corresponding ionized simulations (solid lines) from Papers I and II (from which the ionized gas fraction is also shown as the solid green lines). Note that, in contrast to the corresponding plots in our previous papers, those presented here have a \emph{linear} y--axis.\\
\indent At all epochs, the unbound gas fraction in the ionized calculations is higher than in the corresponding winds runs. In all cases, photoionization unbinds material very quickly in the early phases of the simulations, but the rate at which it unbinds gas tails off rather quickly. This tailing off is, however, not caused by any decreased ability of the ionizing sources to ionize the gas, since the global ionization fractions continue to rise, and in some clouds, the increase in ionization fraction even accelerates somewhat. In some cases, e.g. Run UQ, ionization begins to unbind gas at a faster rate towards the end of the simulations. This is due to the formation of second--generation ionizing sources embedded in the dense bubble walls. The reduced rate at which photoionization expels gas at later stages of the simulations is due to the HII regions bursting out of the clouds and the 10$^{4}$K ionized gas leaking out, reducing the pressure inside the bubbles. This is discussed in more detail later in the paper.\\
\indent By contrast, the winds unbind material at much slower but nearly constant rates. The fractional difference between the quantities of mass unbound by winds and ionization therefore deceases in the later stages of the simulations as the winds begin to catch up.\\
\subsection{Effects of winds on star--formation efficiencies}
\begin{figure*}
     \centering
     \subfloat[Run I]{\includegraphics[width=0.30\textwidth]{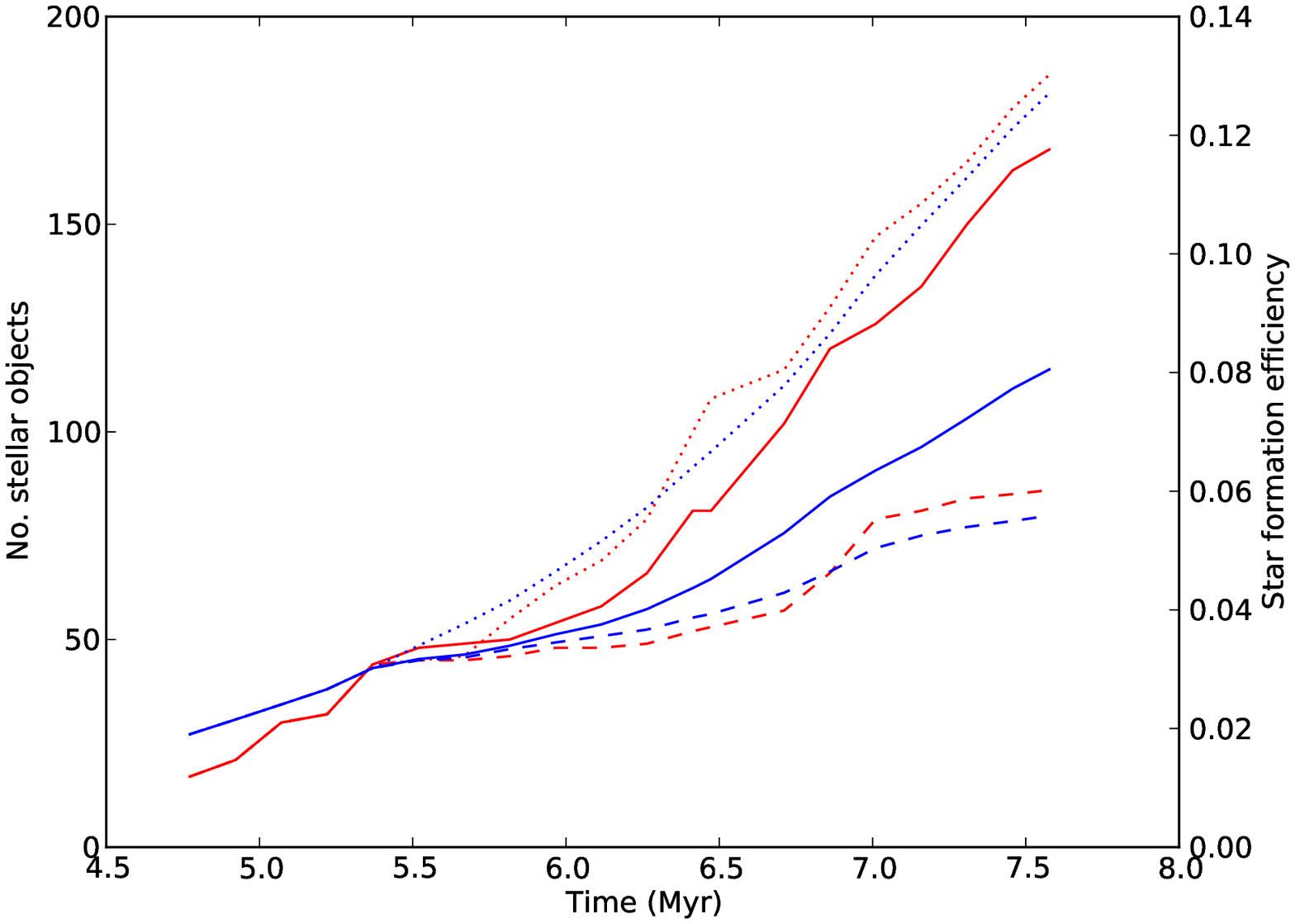}}     
     \hspace{.01in}
     \subfloat[Run J]{\includegraphics[width=0.30\textwidth]{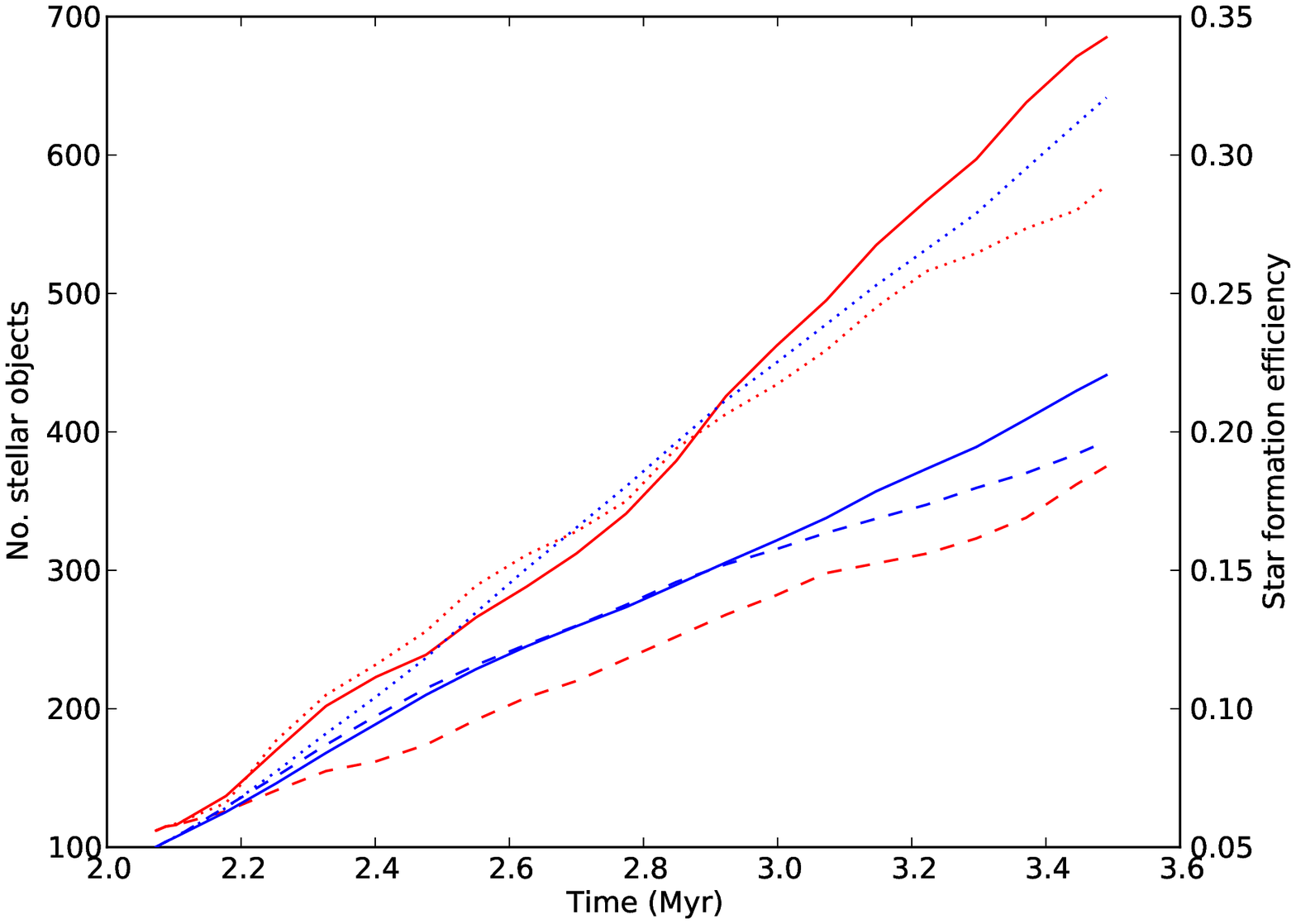}}
     \hspace{.01in}
     \subfloat[Run UP]{\includegraphics[width=0.30\textwidth]{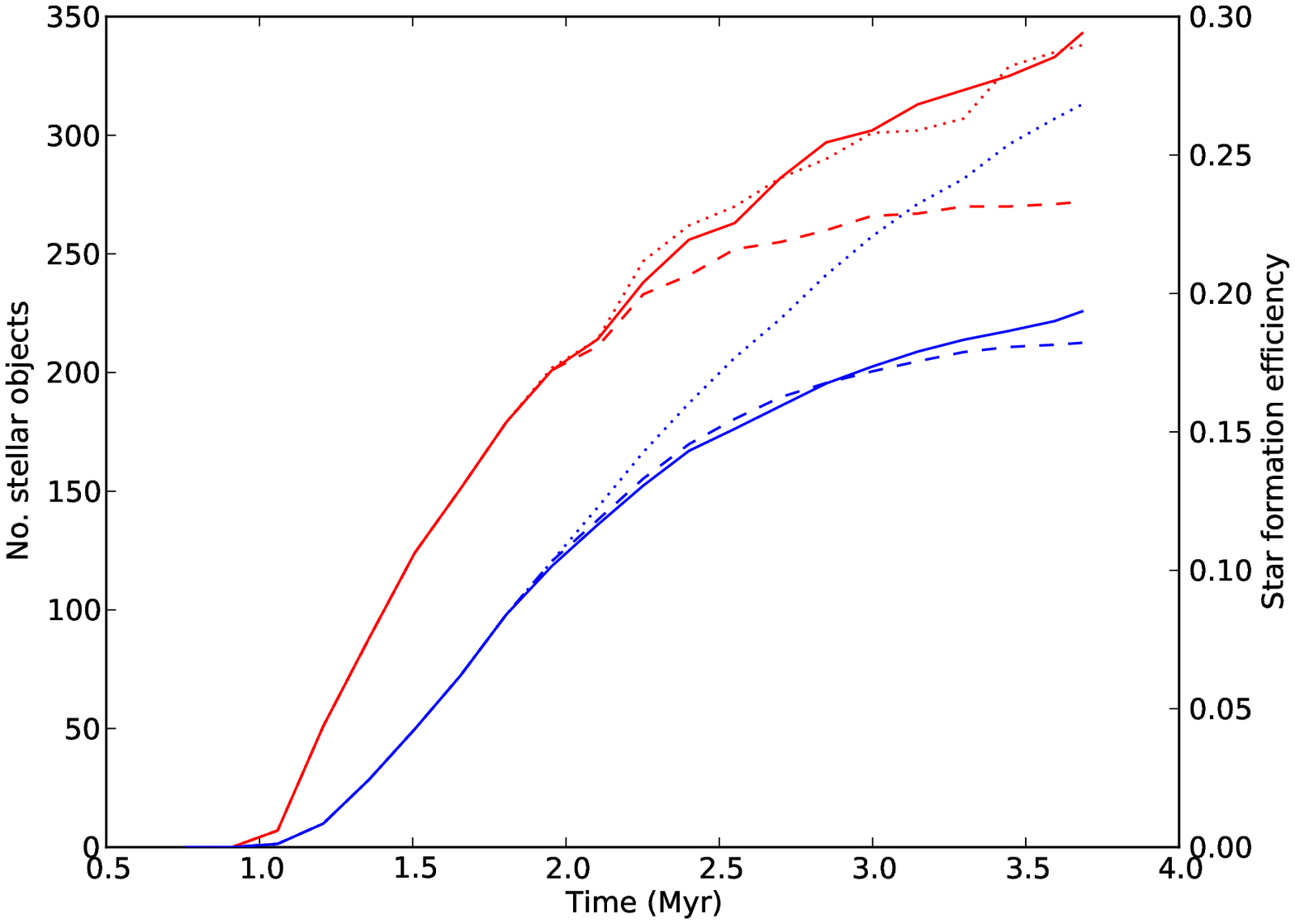}}
     \vspace{.01in}
     \subfloat[Run UV]{\includegraphics[width=0.30\textwidth]{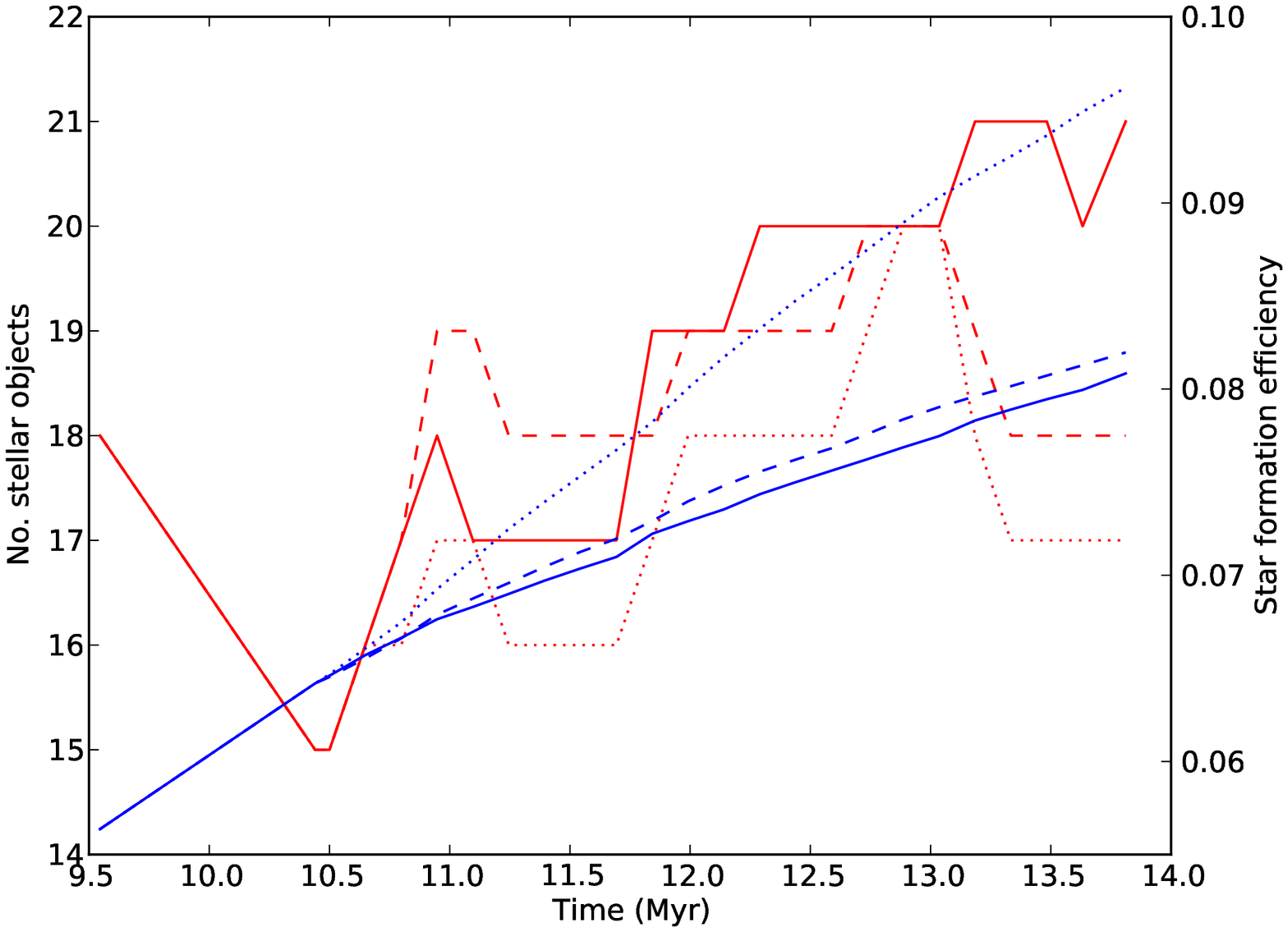}}     
     \hspace{.01in}
     \subfloat[Run UQ]{\includegraphics[width=0.30\textwidth]{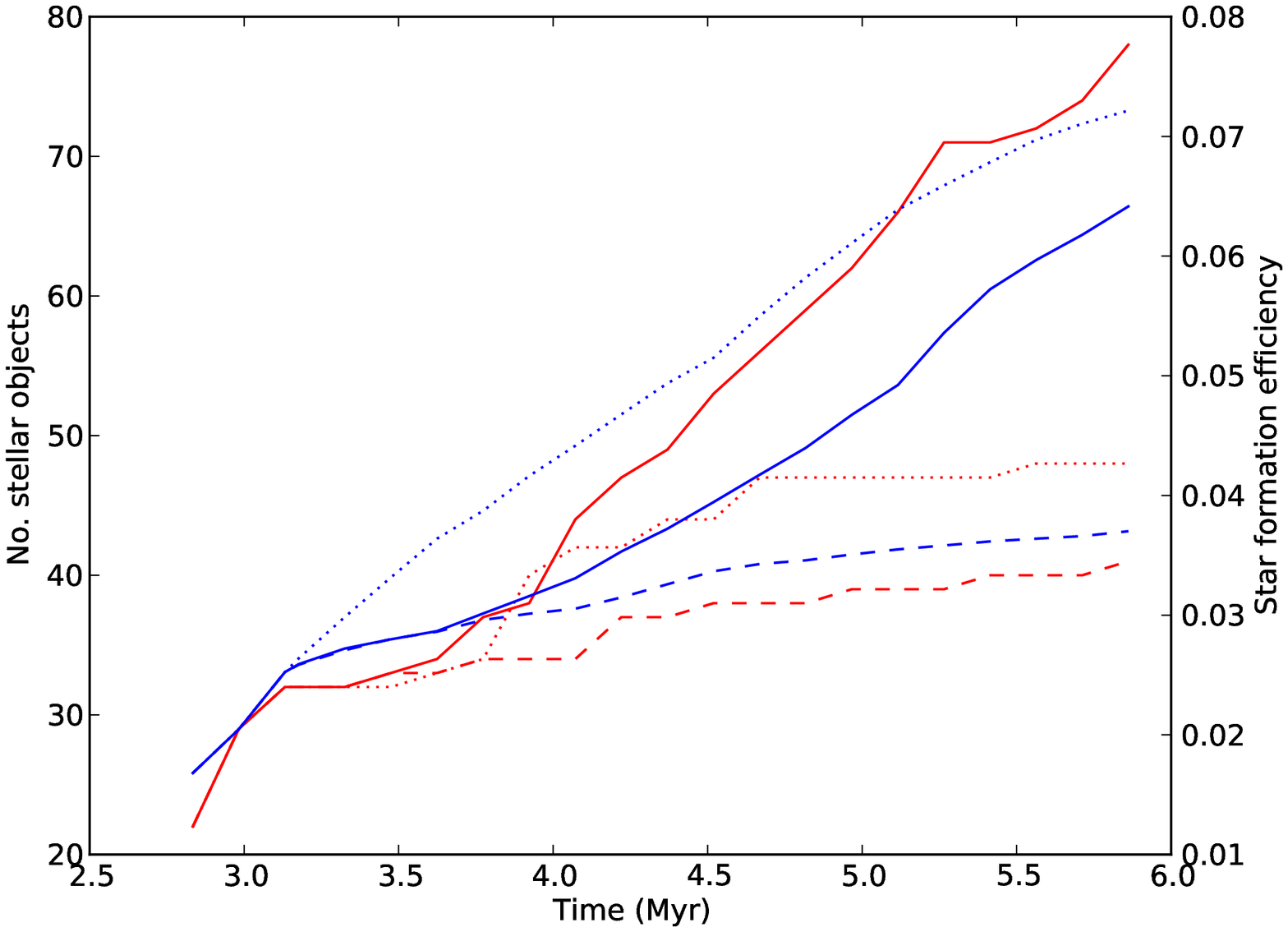}}
     \vspace{.01in}
     \subfloat[Run UF]{\includegraphics[width=0.30\textwidth]{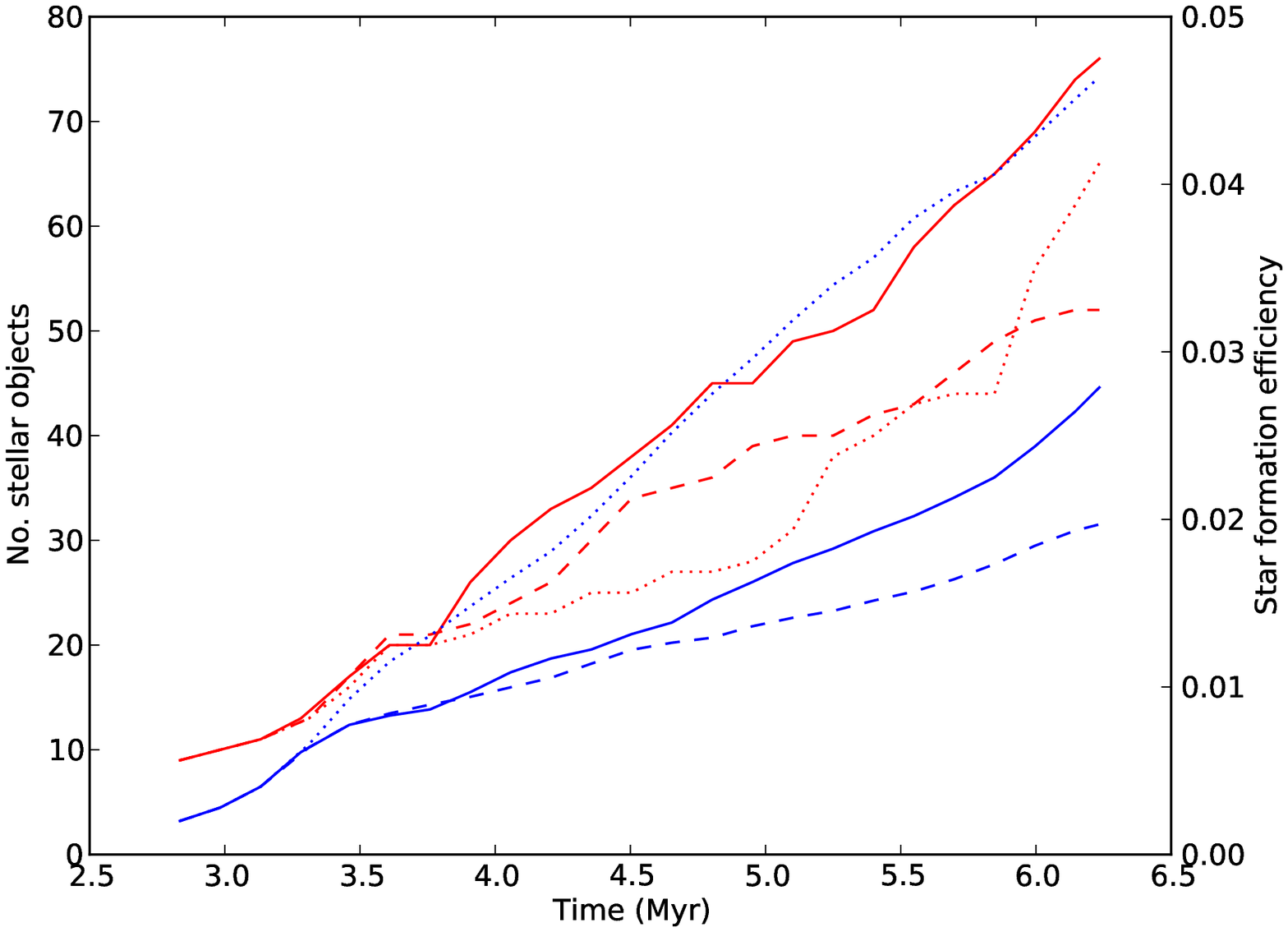}}
\caption{Time variation of the numbers of stellar objects (red lines) and star formation efficiencies (blue lines) in the control (dotted lines), ionized (solid lines) and winds (dashed lines) simulations I, J and UP (top row), and UV, UQ and UF (bottom row).} 
\label{fig:sfe}
\end{figure*}
\begin{figure}
\includegraphics[width=0.5\textwidth]{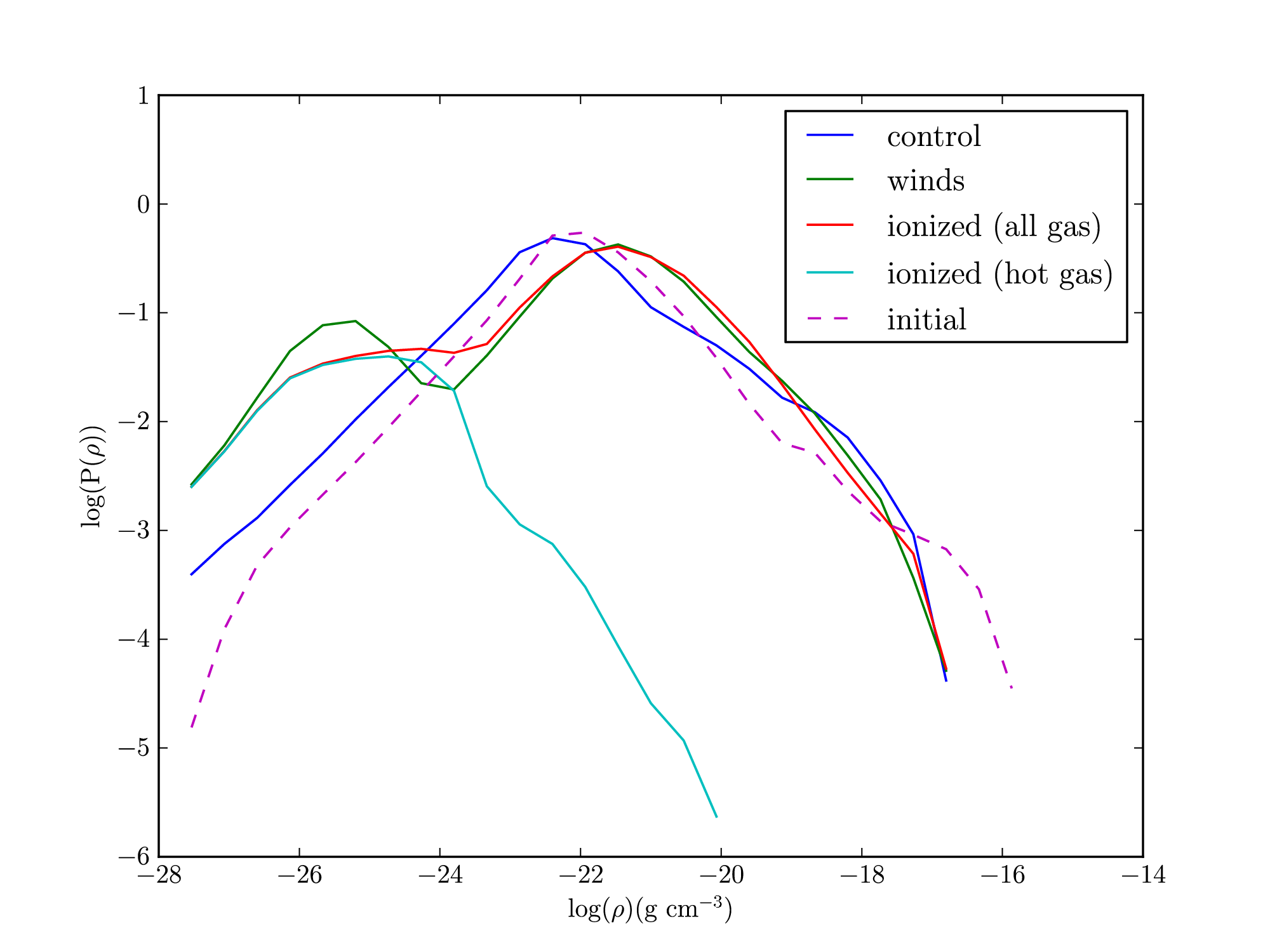}
\caption{Density probability distribution functions for the control (blue line), windblown (green line), ionized, all gas (red line), ionized, HII gas only (cyan line) Runs I at the ends of the simulations after 2.2 Myr of feedback, and for the clouds at the onset of feedback (magenta line).} 
\label{fig:pdf}
\end{figure}
\begin{figure*}
\includegraphics[width=0.95\textwidth]{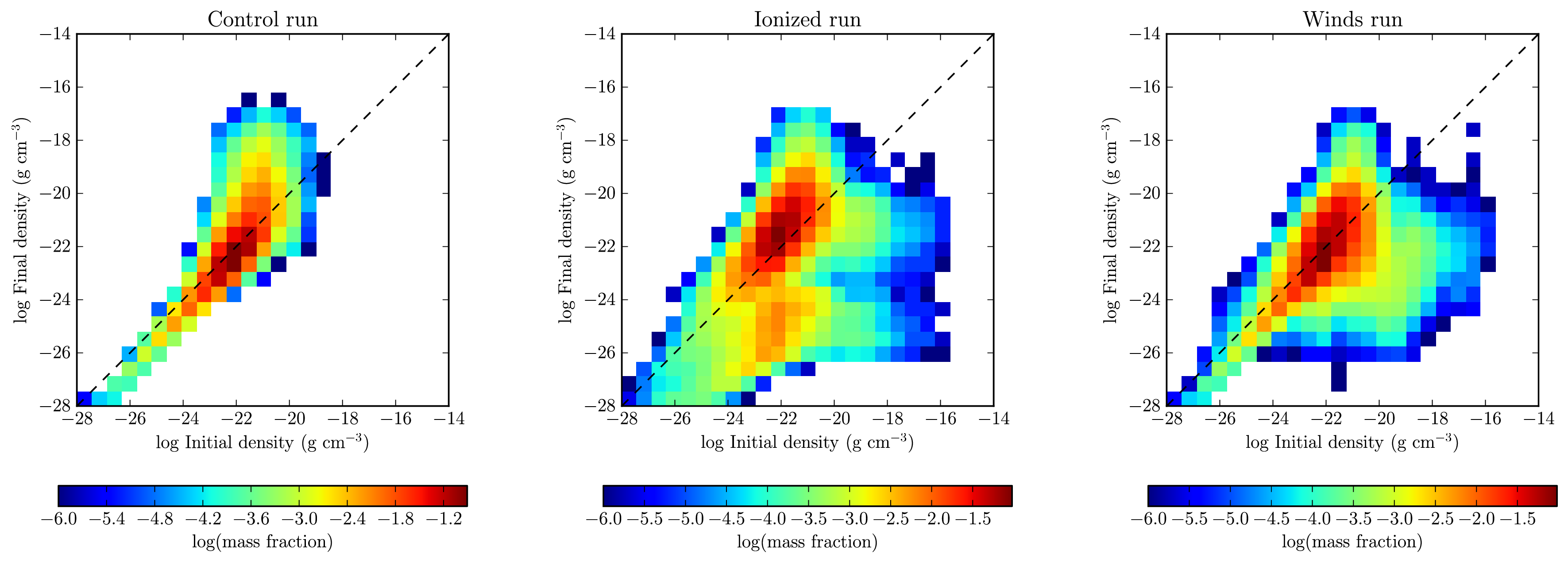}
\caption{Two--dimensional density probability distribution functions for the control (left panel), ionized (centre panel) and windblown (right panel) ionized Runs I. The x--axis in each plot is the gas density at the epoch when feedback was initiated in the ionized and windblown runs, and the y--axis is the gas density at the ends of the simulations after 2.2 Myr of evolution. The black dashed lines indicate the location of gas whose density is unchanged.} 
\label{fig:dvd}
\end{figure*}
\begin{figure}
\includegraphics[width=0.5\textwidth]{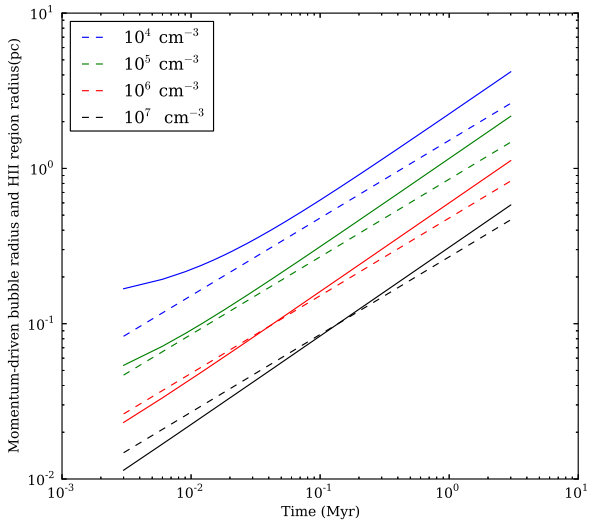}
\caption{Plots of the radius versus time for momentum--driven wind bubbles (dashed lines) and HII regions (solid lines) expanding in uniform media at a range of densities. A wind mass loss rate of 10$^{-6}$M$_{\odot}$ yr$^{-1}$ and terminal velocity of 2 000 km s$^{-1}$, and an ionizing photon flux of 10$^{49}$ s$^{-1}$ have been assumed as representative values.} 
\label{fig:hii_wind_radius}
\end{figure}
\begin{figure*}
     \centering
     \subfloat[Control run]{\includegraphics[width=0.35\textwidth]{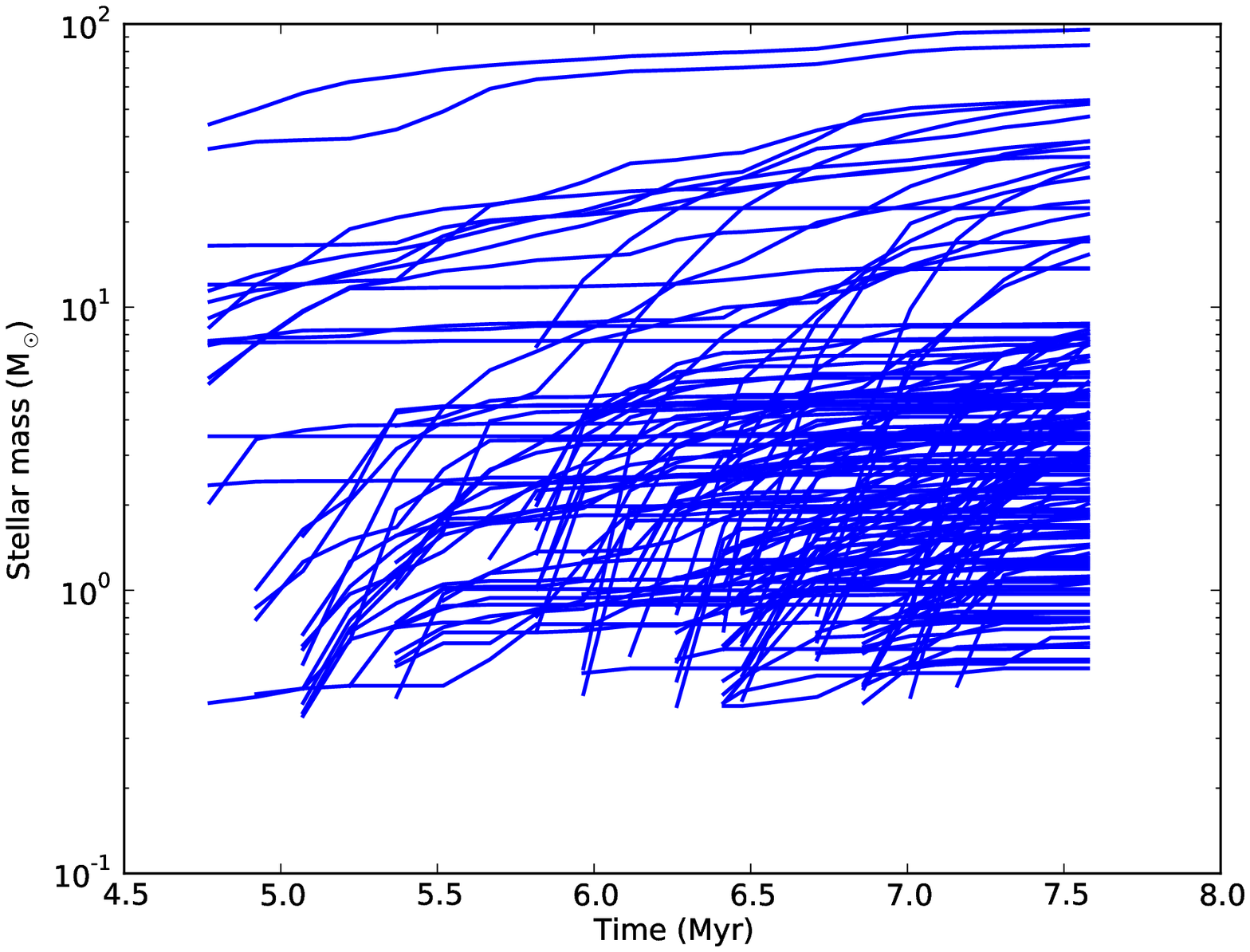}}     
     \hspace{-.25in}
     \subfloat[Ionized run]{\includegraphics[width=0.35\textwidth]{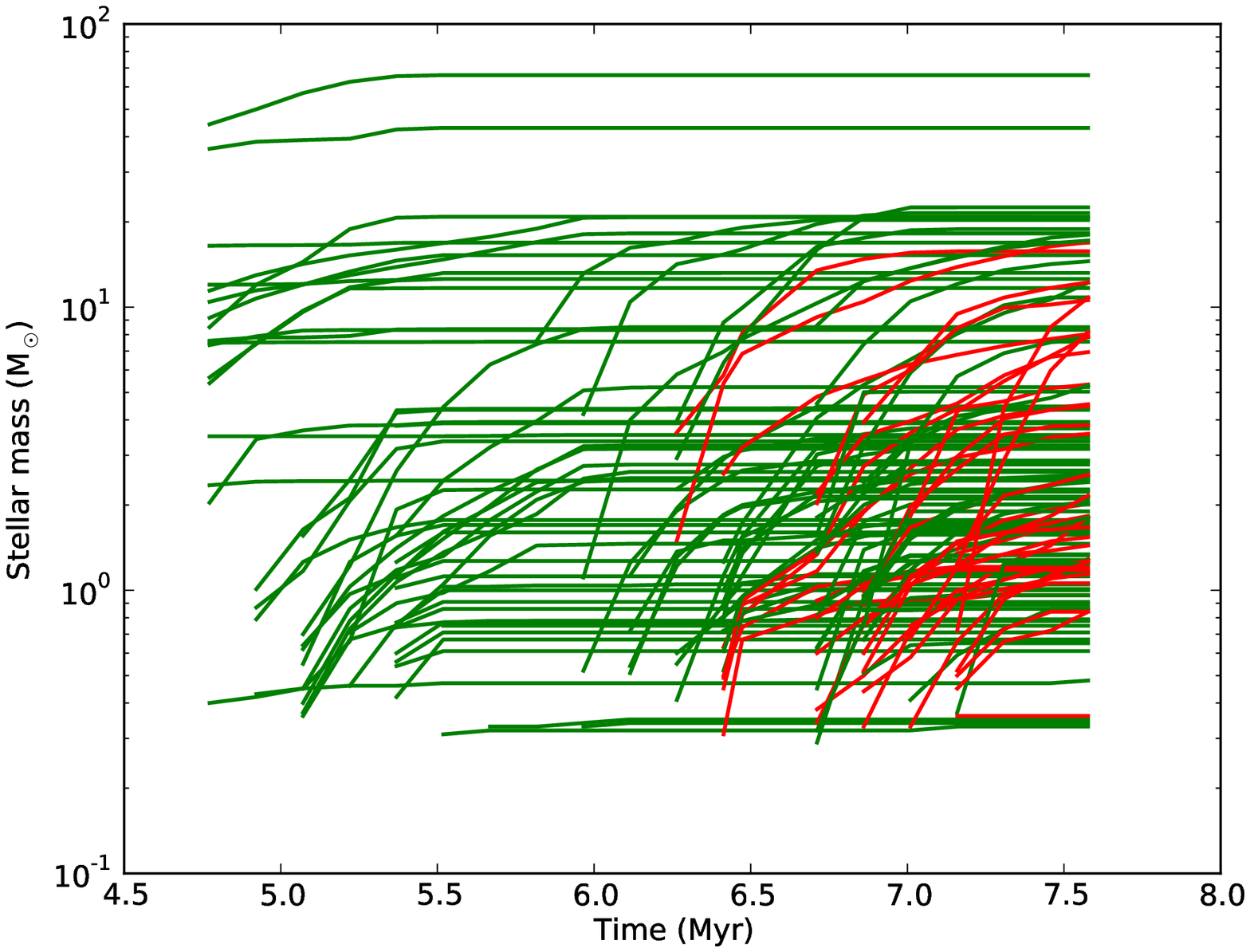}}
     \hspace{-.25in}
     \subfloat[Winds run]{\includegraphics[width=0.35\textwidth]{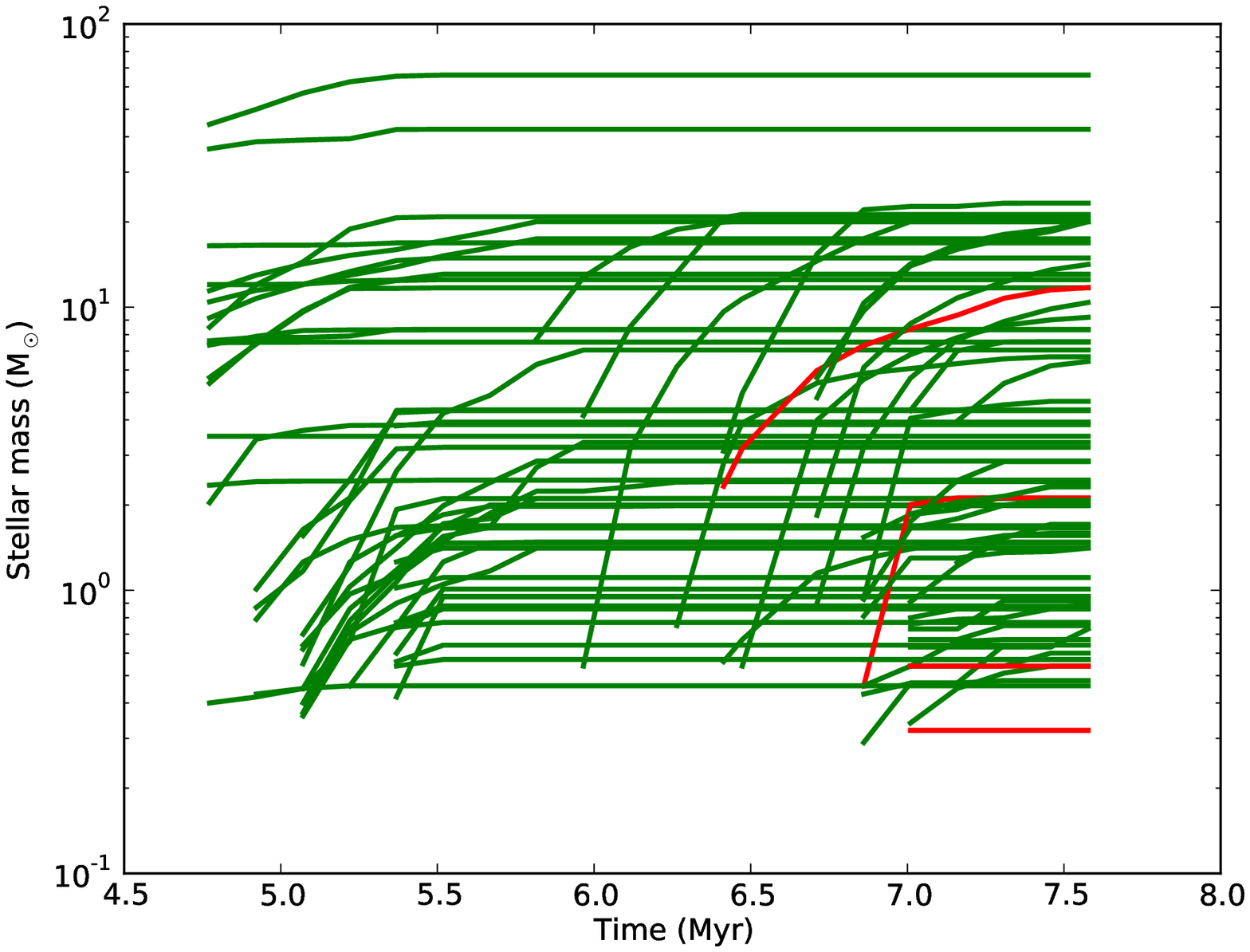}}
\caption{Stellar mass versus time for all stars in the control (left panel), ionized (centre panel) and windblown (right panel) Run I. In the feedback--influenced runs, spontaneously--formed objects are shown in green and triggered objects in red. Note that feedback in these latter runs is initiated at 5.36 Myr.} 
\label{fig:sinkmasses}
\end{figure*}
\begin{figure*}
     \centering
     \subfloat[Control run]{\includegraphics[width=0.35\textwidth]{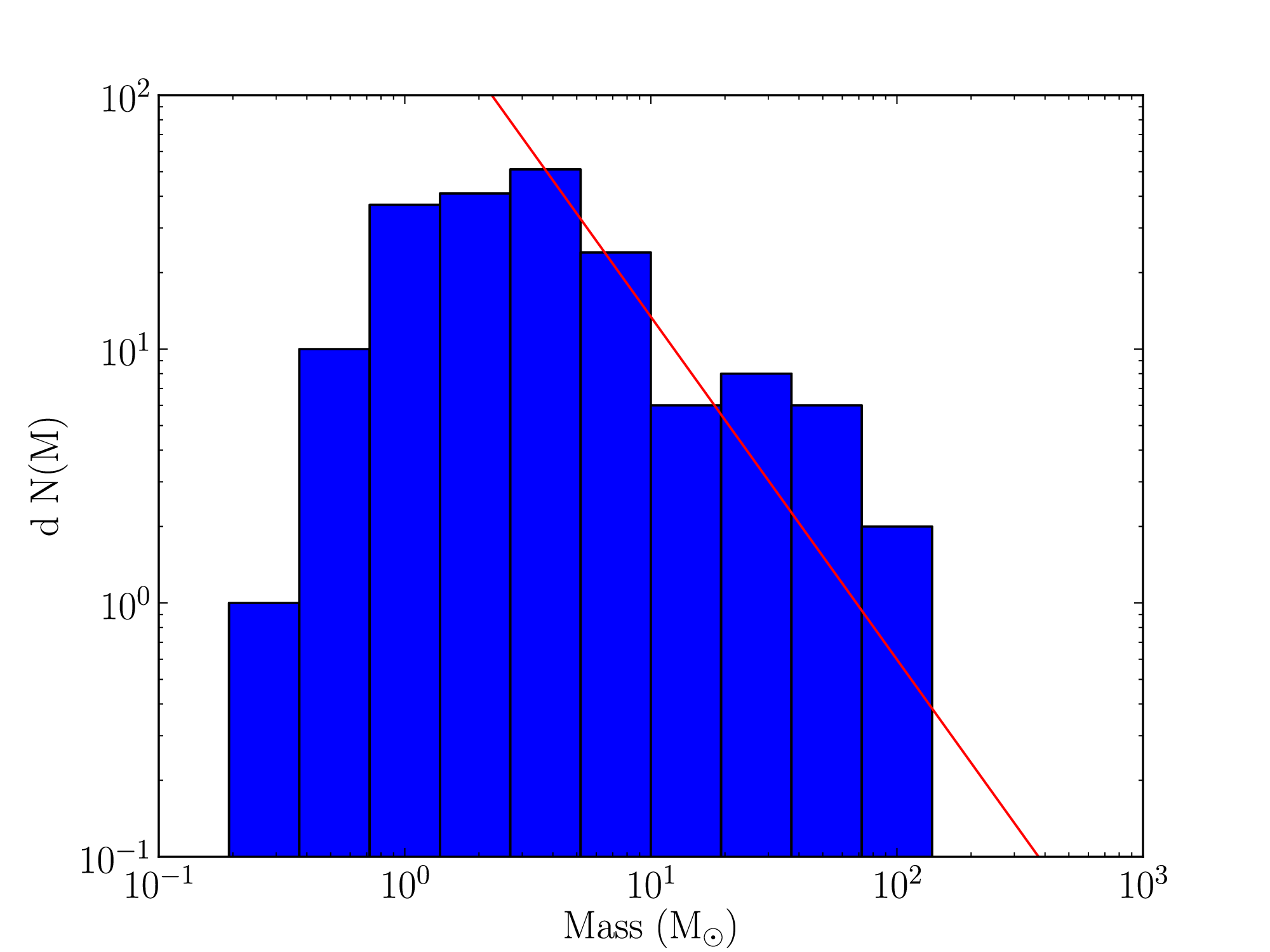}}     
     \hspace{-.25in}
     \subfloat[Ionized run]{\includegraphics[width=0.35\textwidth]{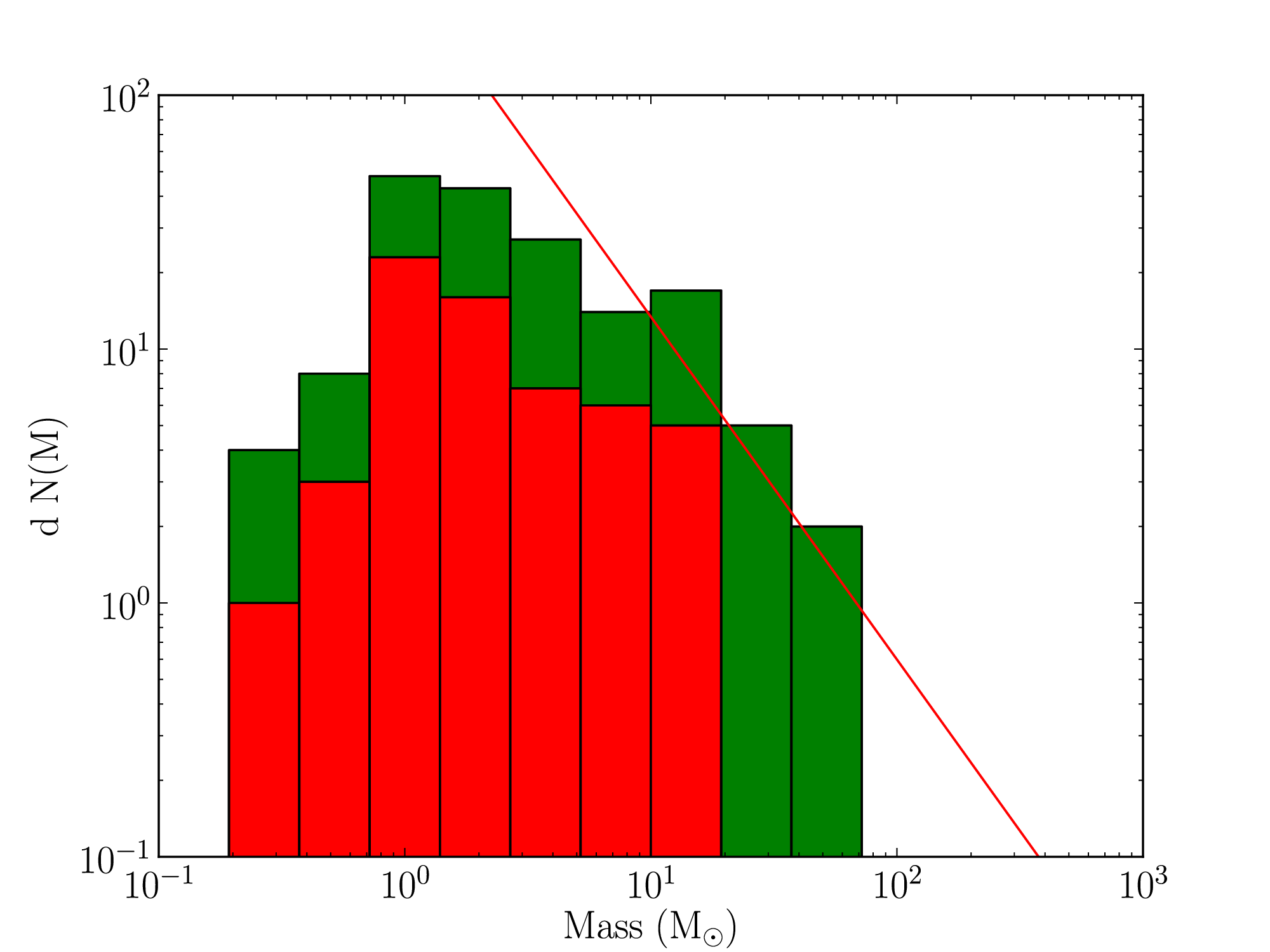}}
     \hspace{-.25in}
     \subfloat[Winds run]{\includegraphics[width=0.35\textwidth]{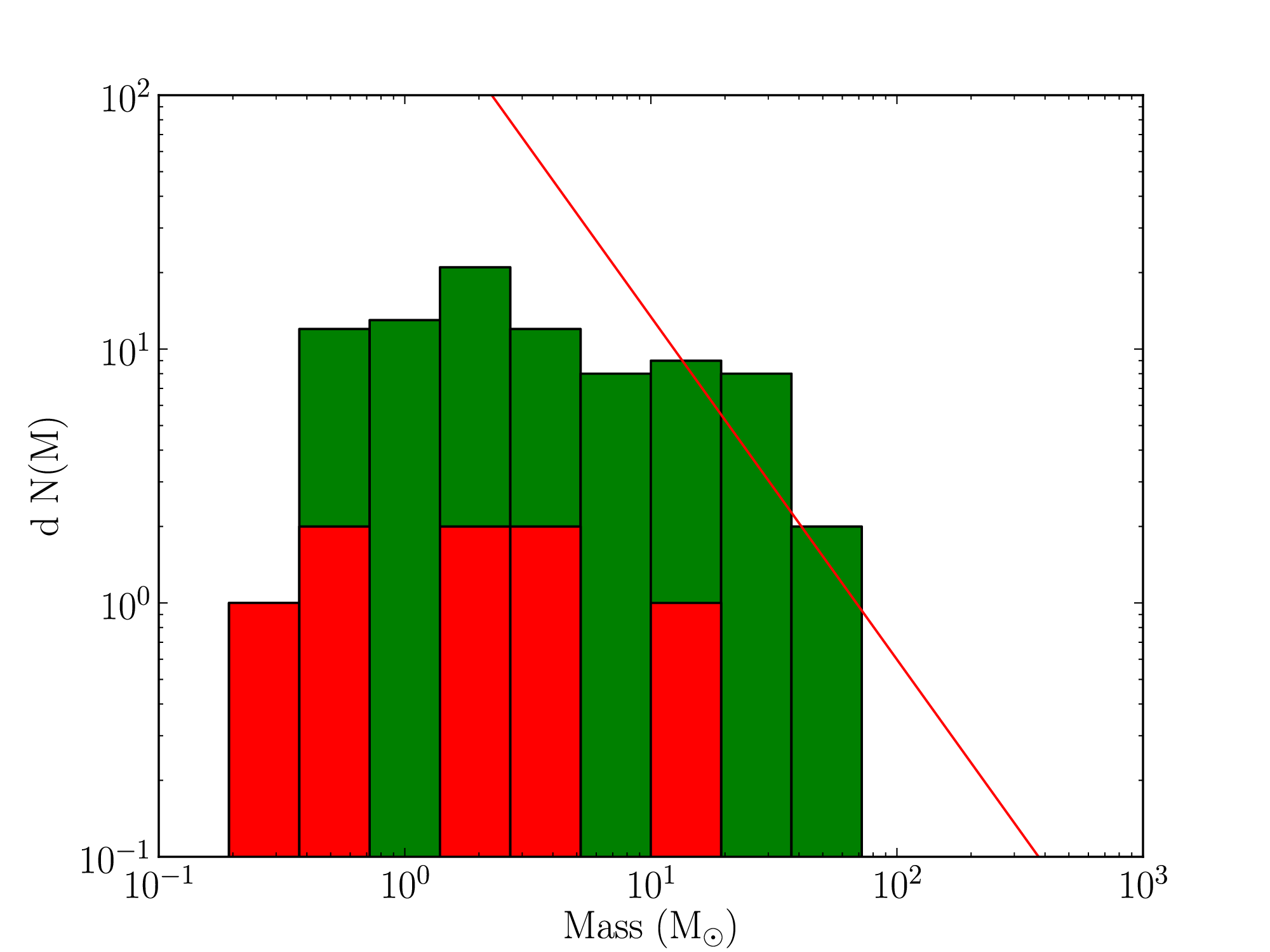}}
\caption{Stellar mass functions at the ends of the control (left panel), ionized (centre panel) and windblown (right panel) Run I. In the feedback--influenced runs, spontaneously--formed objects are shown in green and triggered objects in red. The red line in each plot represents a mass function with a Salpeter slope.} 
\label{fig:massfunc}
\end{figure*}

In Figure \ref{fig:sfe} we plot the variations with time of the numbers of stellar objects and star formation efficiencies in the six selected runs. We compare the ionized (solid lines) and windblown (dashed lines) simulations to the control runs (dotted lines).\\
\indent Photoionization always reduces the star formation efficiency overall relative to the control simulations. In most of the simulations shown here, winds reduce the SFE even further. Except in the case of Run UQ, the difference is rather small however, and these six simulations are those in which winds have the \emph{greatest} dynamical influence on the clouds. In other runs, e.g. Run E or UB, the effect of winds is very small, so that the windblown runs differs little from the control runs. Winds also generally reduce the total numbers of stellar objects born over the timescales under study here relative to both the relevant ionized and control simulations.\\
\indent We used the techniques described in \cite{2012MNRAS.427.2852D} to assess the degree of triggered star formation occurring in the wind--influenced clouds and we found that it was very low, with no more than a few triggered objects being formed in even runs such as I or UQ in which the winds have a substantial influence on the cloud's structure. Momentum--driven winds are evidently much less efficient at inducing star formation than is photoionization.\\
\indent In Figure \ref{fig:pdf}, we show the gas--density probability density function (PDF) for Run I from the time when feedback in the ionized and windblown runs was initiated as a magenta line. The bump at very high densities is a result of a small amount of strongly self--gravitating material which is likely to form one or more sink particles shortly. In general, the PDF extends towards high densities because of self--gravity and towards low densities because of expansion of the cloud, particularly of the very outermost layers.\\
\indent Overplotted are the density PDFs from the windblown run (green line), and two lines from the ionized run showing the PDF of all the gas (red line) and just the ionized gas (cyan line), all from the ends of the simulations after feedback has been acting for $\sim2.2$Myr. The corresponding PDF from the control run after evolving for the same time period without feedback is shown as a blue line.\\
\indent Comparing the evolved control run with the initial conditions, we see that the peak of the PDF has moved to slightly lower densities, due to a small global expansion of the cloud, but that the high--density self--gravitating tail has become much more prominent.\\
\indent The PDF from the ionized run has two peaks. The low--density peak corresponds to the HII region, as shown by the separate cyan line. The peak corresponding to the cold gas ($\sim90\%$ by mass) appears at higher densities than in the control run and there is more gas at intermediate densities ($\sim10^{-21}$--$\sim10^{-19}$ g cm$^{-3}$) in the ionized calculation. However, there is \emph{less} material at very high densities ($>$10$^{-19}$ g cm$^{-3}$) in the ionized run than the control run. These observations are due to the twin effects of ionization discussed in Papers I and II, namely that it destroys the very dense gas in which the O--stars are embedded, but sweeps up lower--density material elsewhere into dense shells. However, because the density profile in the clouds falls off steeply with radius, these dense shells often do not accumulate enough mass to become self-gravitating, and therefore do not attain the extreme densities required to form stars. Hence, the sweeping up of low--density material is not ultimately able to offset or overwhelm the destruction of the high--density star--forming gas. Although there is some triggered star formation in the ionized calculations, the effect on the star formation efficiency is thus uniformly negative.\\
\indent In the windblown clouds, the outcome is similar to that in the ionized clouds, but more extreme. Whereas newly--photoionized gas initially exerts an enormous pressure on its surroundings wherever it may be in the cloud, the ram pressure exerted by the winds falls off very steeply with distance from the massive stars. The winds are thus efficient at destroying the very dense material very close to the sources, or in which the sources are embedded, but they are not as effective as ionization in sweeping up gas elsewhere in the cloud into shells, as shown in Figures \ref{fig:compare_bnd} and \ref{fig:compare_unbnd}.\\
\indent This point can be further illustrated by considering how the densities of each gas particle change during each simulation. In Figure \ref{fig:dvd}, we plot two--dimensional PDFs for the control, ionized and windblown Run I. The x--axis is the gas density at the epoch when feedback is initiated in the two feedback runs. The y--axis is the final density after 2.2Myr of further evolution. The dashed lines show the location of gas whose density remains unchanged.\\
\indent In the case of the control run, the morphology of the plot is quite straightforward. Substantial quantities of gas initially at moderate to high densities becomes denser under the combined influence of self--gravity acting on small scales and continued collision of turbulent flows on larger scales. Meanwhile, most of the gas has changed its density little, since the cloud still enjoys large--scale turbulent support and is not therefore globally collapsing. The low--density tail of the distribution has decreased in density somewhat due to evaporation from the cloud surface, and intermediate and high--density material has also become slightly more rarefied due to gentle global expansion of the cloud.\\
\indent The ionized run is considerably more complex. Again, a significant fraction of the gas has increased in density, partly due to turbulent dissipation and self gravity, as in the control run. However, there is considerably more gas present in this region of the diagram, much of it coming from gas initially at lower densities. This is gas swept up by the expanding HII regions. There are two prominent regions of gas below the y$=$x line. The lower--final--density and brighter region corresponds to the HII region. The higher--initial--density region is material that was originally in the high--density and potentially star--forming central regions of the cloud that was dispersed (but not ionized) by photoionization.\\
\indent Turning to the windblown calculation, we see something intermediate between the two other runs, but closer in form to the ionized run. There is again an enhancement of material raised above the y$=$x contour, extending to very low initial gas densities, but there is not as much of this material as in the ionized run. This indicates that the winds are less efficient at sweeping up initially low-- and intermediate--density material into dense shells, which in turn explains why triggering is less prevalent in the windblown clouds, even when winds have done substantial damage to them.\\
\indent While there is nothing equivalent to the HII region in the windblown PDF, there is clearly substantially more initially intermediate-- and high--density material whose final density is lower. This demonstrates that the winds are more effective at dispersing the initially--dense material near the feeeback sources than are the HII regions. This outcome follows closely the predictions of \cite{2001PASP..113..677C}, who infer that gas number densities must exceed 10$^{5}$ cm$^{-3}$ (i.e. mass densities of $\sim10^{-19}$g cm$^{-3}$) for winds to be more destructive than HII regions. We illustrate this quantitatively in Figure \ref{fig:hii_wind_radius} where we plot the radius with time of expanding momentum--driven wind bubbles (solid lines) and HII regions (dashed lines) in uniform media over a range of densities. We take the wind mass loss rate and terminal velocity to be 10$^{-6}$M$_{\odot}$ yr$^{-1}$ and 2 000 km s$^{-1}$ respectively, and the ionizing photon flux to be 10$^{49}$ s$^{-1}$. At lower densities, the HII region radius is always larger, whereas for the highest densities, the wind bubbles are larger, at least at earlier times. Thus, the densest parts of the clouds in which the O--stars are embedded and where most of the star formation is occurring are more susceptible to wind damage than HII region expansion. This raises the intriguing possibility that, if both winds \emph{and} ionization were active, the winds may assist the expansion of the HII regions by destroying dense gas very close to the massive stars. We will explore this possibility in a subsequent paper.\\
\indent In Figure \ref{fig:sinkmasses} we plot the masses as a function of time of all sinks formed in the control (left panel), ionized (centre panel) and windblown (right panel) Run I calculations. In the latter two plots, spontaneously--formed objects are rendered in green and triggered objects in red. In the feedback runs, many sinks have accretion onto them abruptly terminated, particularly the higher mass objects which are feedback sources, and HII regions and winds are comparable in terms of their ability to halt accretion onto stars. This is of course not the case on the control run where most objects accrete fairly steadily for the duration of the simulation. The figure also shows that the ionized calculation experiences a later burst of star formation beginning at approximately 6 Myr and involving a substantial number of triggered objects. This burst is absent in the windblown calculation, which forms stars quite smoothly and contains very few triggered objects.\\
\indent We also plot in Figure \ref{fig:massfunc} the corresponding mass functions at the end points of all three simulations. The spontaneously--formed and triggered sinks in the feedback--affected runs are again shown in green and red respectively. The mass resolution of these simulations is $\sim0.5-1.0$M$_{\odot}$, so that the low--mass end of the IMF cannot be resolved. However, it can be seen by comparison of these plots that the control and ionized runs produce very similar mass functions whose slopes are roughly consistent with a Salpeter--like power law above 1M$_{\odot}$. Conversely, the windblown calculation produces a much flatter mass function. This is a result of the termination of accretion and star formation in the dense filamentary gas, but without the burst of triggered star formation in the outer regions of the cloud, which produces a substantial fraction of the low--mass stars in the ionized calculation.\\
\indent In summary, winds exert less dynamical influence on a given cloud than HII regions do, and their influence is extremely small for the more massive clouds we have studied. However, when winds are able to strongly influence a given cloud, they are as good as or better than photoionization in dispersing the densest star--forming gas, but much less effective in triggering star formation elsewhere. The influence is essentially confined to smaller spatial scales than that of the HII regions. Their effect on the star formation efficiency, if they are able to influence the clouds' behaviour at all, is thus generally more negative.\\ 
\section{Discussion}
\subsection{Unbinding of gas by winds}
As we saw in Figure \ref{fig:disrupt} above, the fraction of material unbound as a function of time by the winds in any given simulation is always considerably lower than the corresponding unbound masses produced by ionization in the same clouds from \cite{2012MNRAS.424..377D} and \cite{2012arXiv1212.2011D}. However, there are two interesting features of these plots, namely that this deficit often decreases with time and that the rate at which the winds are able to unbind material is approximately linear in time, which we here explain.\\
\indent The radius as a function of time of a wind bubble expanding in a power--law density profile $\rho(r)=\rho_{0}(r/r_{0})^{\alpha}$ may be estimated by equating the rate of change of momentum of the swept--up shell with either the rate of momentum input by the source, or with the force exerted on the shell by the very hot wind gas inside it. These two assumptions are both crude and bracket the true behaviour of the bubble.\\
\indent Using the assumption of momentum conservation,
\begin{eqnarray}
\frac{{\rm d}}{{\rm d}t}(M_{\rm s}v_{\rm s})=\dot{M}v_{\infty},
\label{eqn:ddt}
\end{eqnarray}
where $M_{\rm s}$ and $v_{\rm s}$ are the mass and radial velocity of the swept--up shell and $\dot{M}$ and $v_{\infty}$ are the mass loss rate and terminal velocity of the wind driving the shell's expansion. For simplicity, we will here take the latter two quantities to be constants.\\
\indent If the shell has a radius $R$, we may assume that all the cloud material formerly inside that radius is now contained in the shell and write
\begin{eqnarray}
M_{\rm s}=\int_{0}^{R}4\pi r^{2}\rho(r){\rm d}r=\left[\frac{4\pi\rho_{0}}{(3+\alpha)r_{0}^{\alpha}}R^{3+\alpha}\right]^{R}_{0}.
\end{eqnarray}
For values of $\alpha\geq-2$, the integral does not diverge at $r=0$ and we obtain
\begin{eqnarray}
M_{\rm s}=\frac{4\pi\rho_{0}}{(3+\alpha)r_{0}^{\alpha}}R^{3+\alpha}.
\end{eqnarray}
If we insert this into Equation \ref{eqn:ddt} and make the standard \emph{ansatz} \citep[e.g.][]{1999isw..book.....L} that $R_{\rm MOM}(t)=k_{\rm MOM}t^{n_{\rm MOM}}$, where $k$ is a constant and we have added subscripts to the variables to indicate that they refer to the momentum--conserving solution, we find that 
\begin{eqnarray}
n_{\rm MOM}=2/(4+\alpha)
\end{eqnarray}
and that
\begin{eqnarray}
k_{\rm MOM}=\left[\frac{(4+\alpha)(3+\alpha)r_{0}^{\alpha}\dot{M}v_{\infty}}{8\pi \rho_{0}}\right]^{\frac{1}{4+\alpha}}.
\end{eqnarray}
\indent If we instead assume that the bubble expansion is driven by thermal pressure, the rate of change of momentum of the swept--up shell is equal to the total pressure force acting upon it and 
\begin{eqnarray}
\frac{{\rm d}}{{\rm d}t}(M_{\rm s}v_{\rm s})=4\pi R^{2} P_{\rm w}=4\pi R^{2}\left(\frac{3L_{\rm w}t}{4\pi R^{3}}\right)=\frac{3\dot{M}v_{\infty}^{2}t}{2 R}, 
\end{eqnarray}
where $P_{\rm w}$ is the pressure of the $>10^{6}$K wind gas (assumed to fill the bubble interior) and $L_{\rm w}$ is the mechanical luminosity of the wind, related to the mass loss rate and terminal wind velocity by $L_{\rm w}=(1/2)\dot{M}v_{\infty}^{2}$.\\
\indent Making the same assumptions as before, we get
\begin{eqnarray}
n_{\rm PR}=3/(5+\alpha)
\end{eqnarray}
and that
\begin{eqnarray}
k_{\rm PR}=\left[\frac{3(4+\alpha)(3+\alpha)r_{0}^{\alpha}\dot{M}v_{\infty}^{2}}{16\pi \rho_{0}}\right]^{\frac{1}{5+\alpha}}.
\end{eqnarray}
These results can also be found in \cite{1988RvMP...60....1O}. We see that if $\alpha=0$, these expressions reduce to the standard expansion laws for a uniform medium. Otherwise, the expressions necessarily contain the two arbitrary parameters $\rho_{0}$ and $r_{0}$.\\
\indent Accretion onto a point mass results in a density profile with $\alpha=-3/2$ at small radii \citep{1977ApJ...218..834H}, giving $n_{\rm MOM}=\frac{4}{5}$ and $n_{\rm PR}=\frac{6}{7}$ so, that in such a steep density profile, the expansion power--law indices for the two different wind models are much more similar than they are in a uniform medium.\\
\indent On larger scales and when smoothed over all solid angles, our clouds are well approximated by $\alpha=-2$, as one would expect for an approximately isothermal system. This results in $n_{\rm MOM}=n_{\rm PR}=1$ and  the expansion velocities of bubbles driven by the two mechanisms being constant in time (note that $\alpha=-2$ is the only 3D density profile for which this is the case). The time evolution of momentum-- and pressure--driven wind bubbles in such a density profile are therefore expected to be rather similar.\\
\indent In Figure \ref{fig:vwind}, we plot the mean radial gas velocities at the inner edges of the wind bubbles in the six simulations presented here as functions of time. The bound clouds are shown as solid lines and the unbound clouds as dashed lines. With the exception of Run I, the simulations do indeed exhibit wind bubble expansion velocities that are roughly constant with time as suggested by the above analysis. This is in strong contrast to the behaviour expected in uniform background densities, where the expansion velocity should decline as $t^{-1/2}$. In the case of Run I, the acceleration of the wind bubble is likely due to the fact that this is the cloud most strongly affected by winds and that, in some directions, the bubble rapidly exits the cloud and therefore runs out of braking material to sweep up.\\
Further, if we evaluate $k_{\rm PR}/k_{\rm MOM}$, neglecting small numerical factors, we find that
\begin{eqnarray}
\frac{k_{\rm PR}}{k_{\rm MOM}}\sim\left(\frac{\rho_{0}v_{\infty}}{\dot{M}}\right)^{\frac{1}{6}}r_{0}^{\frac{1}{3}}.
\end{eqnarray}
It is immediately clear that this ratio is very weakly dependent on the relevant variables, but most dependent on the parameter $r_{0}$. Since $n_{\rm MOM}=n_{\rm PR}=1$, we may write 
\begin{eqnarray}
\frac{k_{\rm PR}}{k_{\rm MOM}}=\frac{R_{\rm PR}(t)}{R_{\rm MOM}(t)}=\frac{V_{\rm PR}(t)}{V_{\rm MOM}(t)}\sim\left(\frac{v_{\infty}}{V_{\rm MOM}(t)}\right)^\frac{1}{3}
\end{eqnarray}
From Figure \ref{fig:vwind}, we see that $V_{\rm MOM}(t)\approx7$km s$^{-1}$, so taking $v_{\infty}\sim10^{3}$km s$^{-1}$ gives $k_{\rm PR}/k_{\rm MOM}\approx5$. The difference in normalization of the velocities is therefore significant, but not very large. We explore this issue further in Section 5.4.\\
\indent A consequence of the constant expansion velocity of the wind bubble and the density profile falling as $r^{-2}$ is that the swept--up mass $M_{\rm s}(t)\propto t$, explaining the linear increase with time of the unbound gas masses seen in Figure \ref{fig:disrupt}.\\
\indent In deriving the expressions for $n_{\rm PR}$ and $k_{\rm PR}$ we have assumed that the wind bubble is adiabatic but this is not likely to be true except in the earliest stages of the expansion. Since any loss of energy by radiation or by leakage of the very hot shocked wind gas through holes in the bubble will lead to a decrease in the driving pressure, the expansion of the pressure--driven bubble is actually likely to be slower than linear for most of its evolution. The difference between $k_{\rm PR}$ and $k_{\rm MOM}$ would also be reduced. However, momentum cannot leak or be radiated away, so the linear expansion law of the momentum--driven bubble should be robust. This implies that, in such a density profile, except in the earliest stages of bubble expansion, the bulk of the impulse imparted to the swept--up shell may in fact come from the wind's ram pressure, and not its thermal pressure.\\
\subsection{Dependence of wind damage on cloud properties}
We first note that modelling the winds by injecting momentum only provides a strict lower limit on the dynamical effects of winds. Nevertheless, it seems likely to us that a combination of cooling and the escape of the shocked wind through cavities in the cold gas will greatly lessen the thermal effects of winds. Modelling winds by injecting momentum alone is therefore a reasonable approximation for the time being.\\
\indent In \cite{2012arXiv1212.2011D} we showed that the fraction of material in clouds which could be unbound by photoionization in a fixed common time interval $t_{\rm SN}$ is strongly dependent on the cloud escape velocity. In part this strong correlation was due to the fact that the global ionization fraction of the clouds does not vary very much with cloud properties, and that ionized gas has a fixed sound speed, which regulates its ability to accelerate cold gas by thermal pressure.\\
\indent It is not at all clear that such a simple explanation should hold here. The agent of feedback considered here is the momentum injected by the massive stars. The rate of momentum injection by an individual star is given by the product of the mass--loss rate and the wind terminal velocity, both of which are functions of stellar mass as shown in Figures \ref{fig:mdotfit} and \ref{fig:vinffit}. However, the total rate of momentum input from a population of stars must be computed from an integration over the stellar IMF. If the simplifying assumption is made that all systems have the same IMF, the wind momentum flux is just proportional to the total stellar mass: $\dot{M}_{\rm TOT}v_{\infty}\propto \Sigma M_{*}\propto \eta M_{\rm cloud}$, where $\eta$ is the star--formation efficiency. If the mass loss rate from a fiducial quantity of stellar mass $M_{0}$ is $\dot{M}_{0}$, the momentum injection rate for a given cloud is given by
\begin{eqnarray}
\dot{P}=\dot{M}_{0}v_{\infty},
\end{eqnarray}
assuming a typical value for $v_{\infty}$. We may then evaluate approximately the ability of winds to disrupt clouds by comparing the work done by the winds over the $t_{\rm SN}$ time interval before supernovae detonate to the clouds' binding energies.\\
\indent If the wind bubble expands at a constant velocity $v_{\rm w}$ (note that this is not the same as the wind terminal velocity $v_{\infty}$), the work done by the expansion in the time interval $t_{\rm SN}$ is 
\begin{eqnarray}
E_{\rm w}=\dot{M}_{\rm TOT}v_{\infty}v_{\rm w}t_{\rm SN}.
\end{eqnarray}
The cloud binding energy may be written in terms of the escape velocity approximately as $E_{\rm bind}\approx M_{\rm cloud}v_{\rm ESC}^{2}$, so that the ratio of input wind energy to the cloud binding energy, which we take as an estimate of the unbound mass fraction, is
\begin{eqnarray}
f_{\rm s}=\left(\frac{\dot{M}_{\rm TOT}t_{\rm SN}}{M_{\rm cloud}}\right)\left(\frac{v_{\rm w}v_{\infty}}{v_{\rm ESC}^{2}}\right)
\label{eqn:unb}
\end{eqnarray}
We have assumed here that the cloud binding energy does not change intrinsically over the timescales upon which the winds are active. Contraction of the clouds is indeed insignificant over these timescales. Although some clouds have freefall times comparable to or shorter than the 3Myr feedback timescale, they are not actually in free fall becuase they still enjoy substantial turbulent support. We examined the evolution of the control runs and found that, in the absence of any kinds of perturbation, the change in gravitational binding energy of the clouds during the duration of the simulations was less than 20$\%$. The exception was Run UP, which has the shortest freefall time. Even for this run, the change in gravitational binding energy in the control run was less than a factor of two. In any case, given the simplicity of the argument presented above, we think it is reasonable to neglect intrinsic changes in the clouds' binding energies.\\
\indent We again recover a strong dependence on the cloud escape velocity, but note that it must be compared with two other velocities -- $v_{\infty}$ which is a property intrinsic to OB stars and is therefore known in advance, and $v_{\rm w}$, the expansion rate of the wind bubble, which depends on the properties of individual clouds. The first term in the equation, the fraction of mass returned to the clouds by the winds, can be estimated as $10^{-4}$--$10^{-3}$, since a $10^{4}$M$_{\odot}$ cloud is likely to host a few O--stars with wind fluxes of $\sim10^{-6}$M$_{\odot}$ yr$^{-1}$.\\
\begin{figure}
\includegraphics[width=0.5\textwidth]{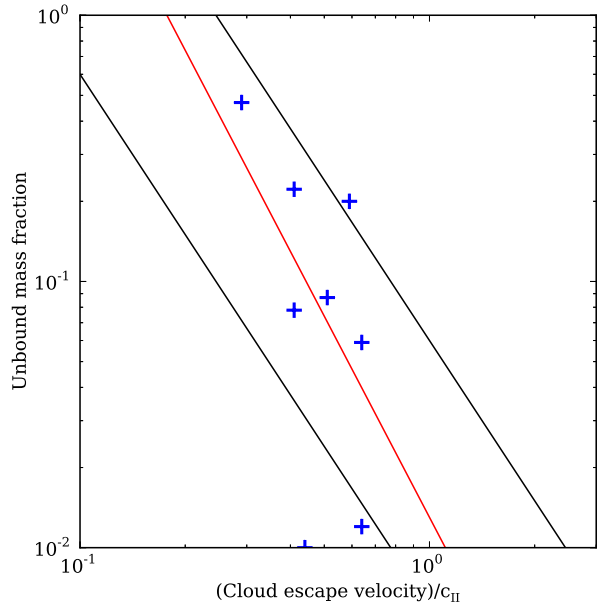}
\caption{Plot of fractions of mass unbound by winds in clouds against cloud escape velocities, normalised by the ionized sound speed for ease of comparison with Figure 8 in Dale et al 2012c. Blue crosses are simulation results, the red line is a fit to the simulation results, and the black lines are solutions to Equation 14 assuming values for $v_{\rm w}$ of 6 kms$^{-1}$ and $\dot{M}t_{\rm SN}/M_{\rm cloud}$ of 10$^{-3}$ (upper line) and 10$^{-4}$ (lower line).} 
\label{fig:unb_vesc}
\end{figure}
\indent In Figure \ref{fig:unb_vesc}, we plot the fractions of gas unbound in all the windblown simulations presented here as a function of the clouds' escape velocities. The red line is a least--squares fit to the data with a logarithmic slope of -2.5. The black lines express Equation \ref{eqn:unb} assuming returned mass fractions of $10^{-4}$ and $10^{-3}$ and assuming a value of 6 km s$^{-1}$ for $v_{\rm w}$, a typical value from Figure \ref{fig:vwind} above. There is clearly a great deal of scatter about the fit. Although most of the points lie between the black lines described by Equation \ref{eqn:unb}, the most that can be said is that the results are not inconsistent with the analysis presented above, and that the unbound gas fractions drop very steeply as the cloud mass and escape velocity increase. Comparison with Figure 8 from \cite{2012arXiv1212.2011D} shows that the largest unbound gas fractions achieved by winds are comparable to, albeit lower than, those reached in the ionized models.\\
\begin{figure}
\includegraphics[width=0.5\textwidth]{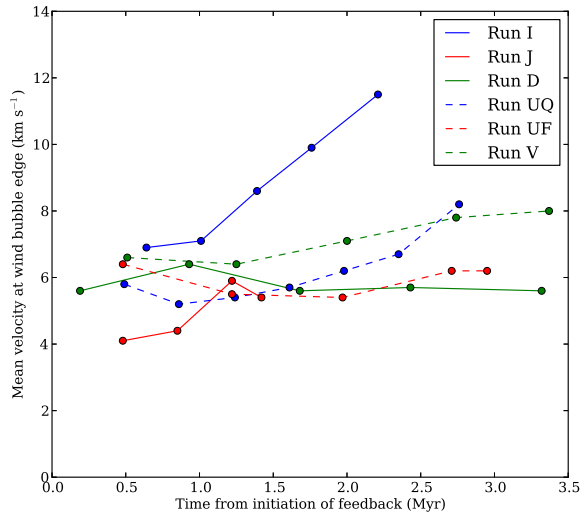}
\caption{Plot of mean radial gas velocities at the inner edge of the wind bubbles (as measured by the velocities of the working--face particles) against time for six windblown simulations, showing that, for most calculations, the expansion velocity of the bubbles is close to constant.} 
\label{fig:vwind}
\end{figure}
\subsection{Inability of winds to trigger star formation}
\begin{figure}
\includegraphics[width=0.5\textwidth]{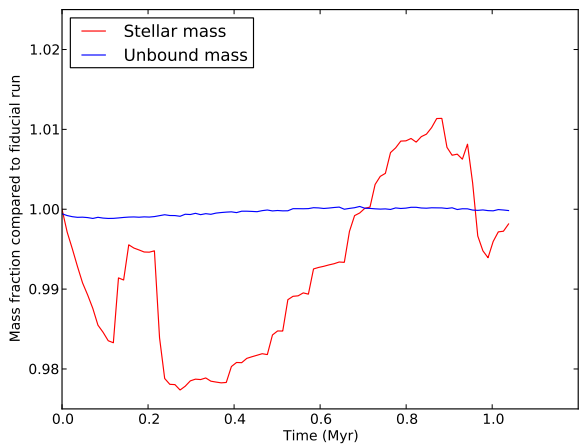}
\caption{A comparison of the unbound mass (blue line) and star formation efficiency (red line) in the high--resolution and fiducial runs UB (line represent values in the high--resolution run divided by values in the low--resolution run).} 
\label{fig:runa_res_plot}
\end{figure}
\begin{figure}
\includegraphics[width=0.55\textwidth]{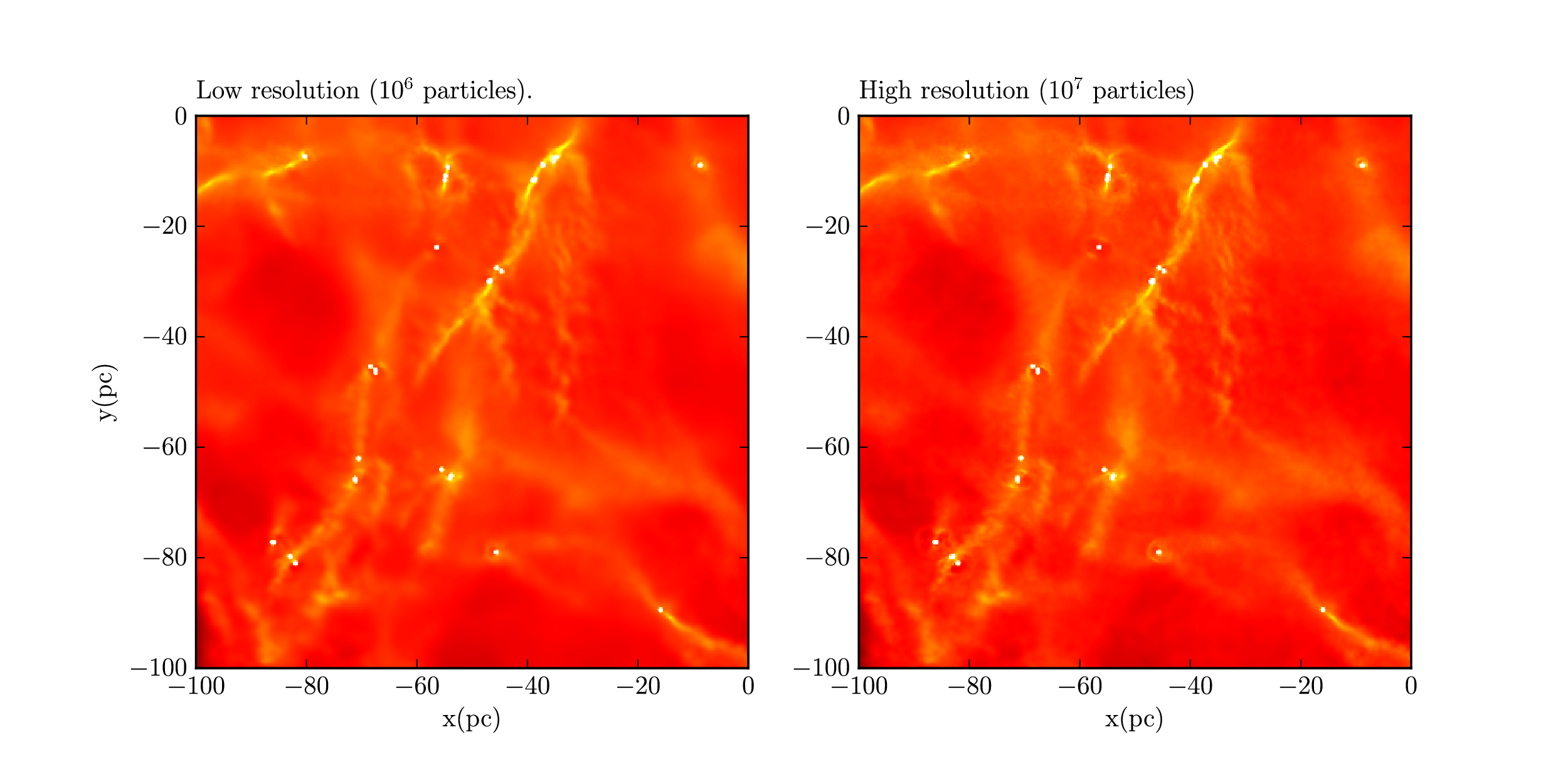}
\caption{Column--density plots of the fiducial (left) and high--resolution (right) runs UB after $\sim1$Myr of evolution. The images have been zoomed in to show the main region of star formation.} 
\label{fig:runa_res_snap}
\end{figure}
\indent Winds appear to have little ability to trigger star formation in our calculations. To check that this is not a resolution effect, we repeated Runs A, B and UB with ten times more particles (that is, 10$^{7}$ instead of 10$^{6}$). This was achieved by taking the simulation output dump at the onset of feedback and, for each active SPH particle, adding nine new particles at random locations within its smoothing kernel with the same velocities as the parent particle, then renormalizing the particles masses. This makes the simulations time--consuming, so we were not able to run them for the canonical 3Myr. We show in Figure \ref{fig:runa_res_plot} where we compare the unbound mass fraction and star formation efficiency as functions of time between the original and higher--resolution runs UB. We plot the values for the high--resolution calculation divided by the values for the standard calculation. The maximum deviation in the star formation efficiency over the $\sim1$ Myr period observed is less than 1.5\%, and in the unbound mass fraction is less than 0.5\%. These figures are representative of the other high--resolution comparison runs. Figure \ref{fig:runa_res_snap} shows snapshots, this time from the Run A comparison, after $\sim0.7$ Myr of evolution, which are very nearly indistinguishable. We conclude, therefore, that this result is not a resolution issue.\\
\indent There are two widely--used scenarios for the triggering of star formation -- the collect--and--collapse model \citep[e.g][]{1978ApJ...220.1051E} and the cloud--crushing model \citep[e.g.][]{1982ApJ...260..183S}. It seems that neither operates efficiently in our simulations.\\
\indent The cloud--crushing model relies on the over--running by shocks of stable or quasi--stable globules of material which are significantly denser than the background and has been extensively studied by many authors \citep[e.g.][]{1982ApJ...260..183S,1984ApJ...282..178S,1994A&A...289..559L,2006ApJS..164..477N}. This model is a strong example of triggered star formation, since the initial conditions are stable by construction and it is only the external influence of radiation or shocks which initiates star formation. However, it is difficult to see how the model can be applied in the clouds studied here because most of the inhomogeneities in the gas are linear or filamentary structures. There are no isolated rounded structures which are likely to suffer shock--driven implosion. The massive stars are almost exclusively born inside these filaments, often at junctions where several meet (as observed by \cite{2012A&A...540L..11S}).\\
\indent From the perspective of the feedback sources, the dense filaments are mostly accretion flows and the ability of feedback to affect their evolution depends on the ram pressure in the flows compared with the ram pressure exerted on them by the wind sources. The typical gas densities in the filamentary structures observed in our calculations have typical values of $10^{3}$--$10^{4}$ cm$^{-3}$, with the less massive clouds tending to have the lower density filaments. The flow velocities inside the filaments in the $10^{4}$M$_{\odot}$ clouds are 1--2 km s$^{-1}$, whereas they are 3--5 km s$^{-1}$ in the more massive clouds due to their higher turbulent velocities and also to their deeper potentials. Ram pressures in the filamentary flows are substantially higher in the more massive clouds. Individual clusters in the low-- and high--mass clouds tend to have similar total momentum fluxes however -- typical mass fluxes are $\dot{M}_{\rm cluster}\sim10^{-5}$M$_{\odot}$yr$^{-1}$ and wind velocities are 2 000 km s$^{-1}$. The outward ram pressure from the cluster winds drops with radius $r_{\rm clus}$ from each cluster as $r_{\rm clus}^{-2}$, so the value of $r_{\rm clus}$ at which the wind ram pressures and the accretion flow ram pressures become equal can be estimated, yielding several pc in the low--mass clouds, but less than one pc in the higher--mass clouds.\\
\indent This implies that winds are able to disrupt and reverse the accretion flows before they penetrate the clusters in the lower--mass clouds, but not in the high--mass clouds (typical cluster radii in the low--mass calculations are $\sim$1pc and we have assumed that the clusters in the high--mass runs, which are represented by single sink particles, have similar sizes). Figures \ref{fig:compare_bnd}, \ref{fig:compare_unbnd} and \ref{fig:runa_res_snap} confirm this picture and show clearly that most of the structures generated by the winds are perpendicular to the filaments and expanding into the lower--density ambient gas. In any case, the ability to disrupt the accretion flows results in \emph{less} star formation and not more, since preventing the accretion flows from entering the clusters outweighs any triggering affects in the eroded tips of the flows.\\
\indent Turning to the collect--and--collapse model, \cite{1994ApJ...427..384E} presented a straightforward technique for computing the growth rate of instabilities in expanding shells. If, instead of a uniform background medium, a medium characterised by $\rho(r)=\rho_{0}(r/r_{0})^{-2}$ is assumed, the dispersion relation for perturbations on a spherical shell with angular wavenumber $\eta$ is
\begin{eqnarray}
\omega(\eta)=\frac{-2V(t)}{R(t)}+\left(\frac{2\pi G\rho_{0}r_{0}^{2}\eta}{R(t)^{2}}-\frac{c_{\rm s}^{2}\eta^{2}}{R(t)^{2}}\right)^{\frac{1}{2}}
\label{eqn:disp}
\end{eqnarray}
If $V(t)=$constant$V_{0}$ (which we take to be 7 km s$^{-1}$ once again), $R(t)=V_{0}t$ and $\omega(\eta)$ can be evaluated if $\rho_{0}r_{0}^{2}$ is known. Since our model clouds have roughly constant column--density, $M\propto R^{2}$ and $\rho_{0}r_{0}^{2}$ therefore scales with $R$, so the first term inside the square root in Equation \ref{eqn:disp} should be largest for the largest clouds, all else being equal. To measure the total degree of fragmentation expected as a function of $\eta$, we may compute the fragmentation integral \citep{2009MNRAS.398.1537D}
\begin{eqnarray}
I_{\rm f}(\eta,t_{1})=\int_{t_{0}}^{t_{1}}\omega(\eta,t){\rm d}t
\end{eqnarray}
Evidently, $t_{0}$ cannot be set to zero because all the terms in the dispersion relation diverge at this point. Additionally, at very early times when $R(t)$ is small, $\omega$ may well be negative, leading to a negative fragmentation integral. This is not physical, so we set any values of $\omega$ less than zero to zero to obtain an upper limit on the fragmentation integral. We set $t_{0}$ to a small value of 10$^{4}$ yr and $t_{1}$ to 3Myr. We then evaluated the fragmentation integrals for the simulations discussed here, measuring values of $\rho_{0}$ and $r_{0}$ from the density profiles of the clouds. We found that, for all except a few of the largest clouds (e.g. A and UV), the fragmentation integral was zero for all wavenumbers, and even for the largest clouds, had a maximum value of only $\approx$2.\\
\indent The reasons for the lack of instability in shells expanding into an $r^{-2}$ density profile are twofold. The first term in the dispersion relation is a stabilizing term which results from the transverse spreading of perturbations by the expansion of the shell, against which the self-gravity of the perturbations must fight. A shell expanding into a density profile shallower than $r^{-2}$ decelerates, so that the stabilizing term decreases both due to $V(t)$ becoming smaller and $R(t)$ becoming larger. As detailed above, in the clouds studied here, $V(t)$ is constant, so the stabilizing term shrinks more slowly. Secondly, the first term inside the brackets in the dispersion relation derives from the self--gravity of perturbations on the shell. The form of this term is very different to that obtaining in a uniform background medium, being proportional to $R^{-2}$ instead of being a constant for a given wavenumber. This in turn is due to the different evolution of the shell surface density. In a uniform background medium, the surface density of an expanding shell increases with time as $R$, but in a $r^{-2}$ density profile, the surface density decreases as $R^{-1}$. It is indeed unlikely that the collect--and--collapse process would operate in our model clouds.\\
\subsection{Relative effects of HII regions and winds}
\begin{figure}
\includegraphics[width=0.5\textwidth]{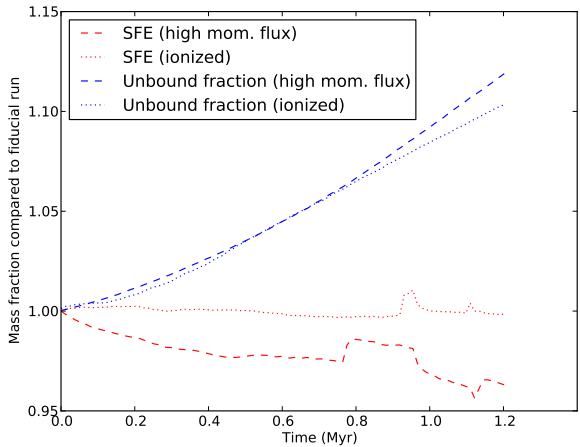}
\caption{A comparison of the unbound mass (blue lines) and star formation efficiency (red lines) in the enhanced momentum flux (dashed lines) and ionized (dotted lines) runs UB compared to the fiducial UB wind run.} 
\label{fig:ubclx25_plot}
\end{figure}

\begin{figure*}
\centering
\includegraphics[width=1.15\textwidth]{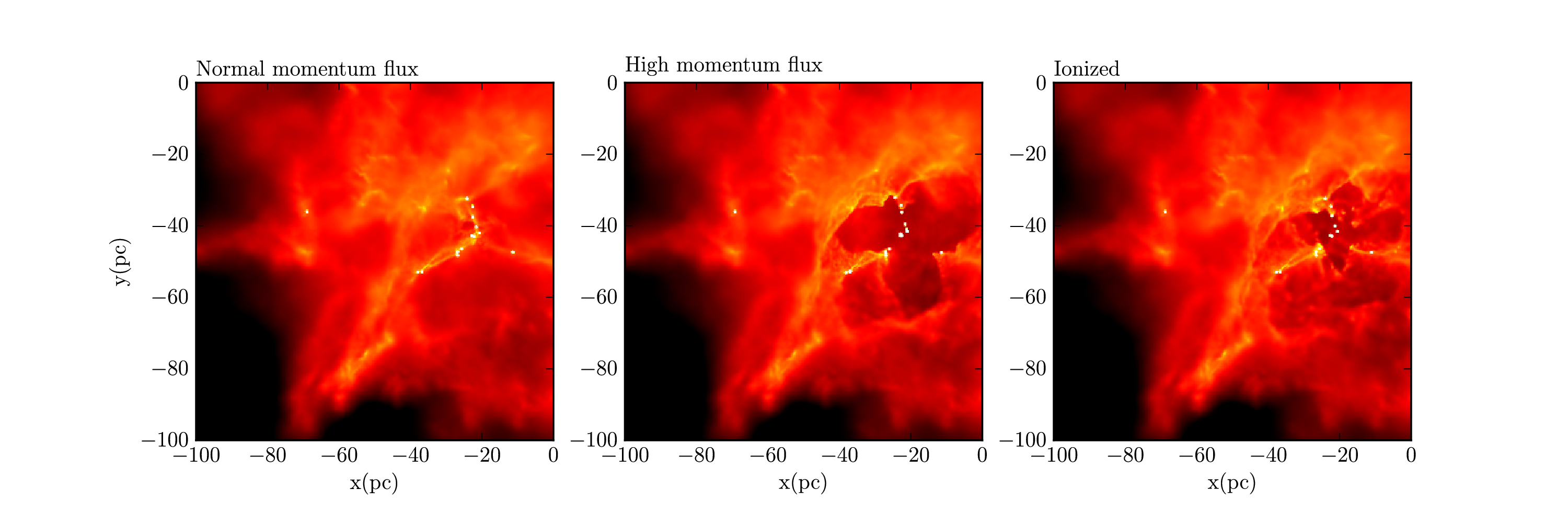}
\caption{Column--density snapshots from the enhanced momentum flux (centre panel) and ionized (right panel) runs UB compared to the fiducial UB wind run (left panel).} 
\label{fig:ubclx25_snap}
\end{figure*}

\begin{figure}
\includegraphics[width=0.5\textwidth]{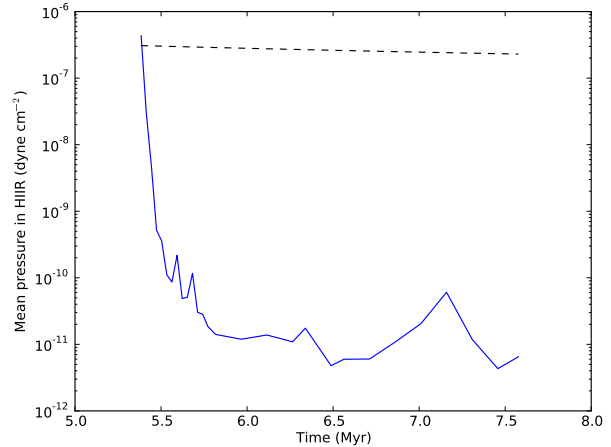}
\caption{The evolution with time of the mean pressure in the HII region in Run I (blue line) compared to the evolution of pressure in an HII region expanding in a uniform background (dashed black line).} 
\label{fig:HII_press}
\end{figure}

\indent A similar but more sophisticated analysis to that outlined above was performed for the case of HII regions expanding in smooth power--law density profiles by \cite{1990ApJ...349..126F}. They showed that, if $\alpha < -3/2$, such an expansion is in fact unstable because the rate at which the expanding ionized bubble sweeps up mass is insufficient to prevent the swept--up shell itself being eventually ionized, at which point the HII region bursts through it into the cloud beyond. This implies that, since our model clouds have average density profiles steeper than this, ionization would have rapidly and completely ionized them if they were smooth. That this does not occur is due to several factors. The strong inhomogeneities in the gas resulting from the combined effects of turbulence and gravity protect the clouds by ensuring that the radiation from a given source is not able to overrun the entire cloud. In addition, the accretion flows onto the ionizing sources and the clouds' gravitational fields (which were neglected by \cite{1990ApJ...349..126F}) both retard the ability of the HII regions to expand, at least in some directions.\\
\indent However, from a given source, it is likely that there will be at least some directions in which the radial density profile is smooth and steep. As the HII regions expand in these directions, the material they sweep up will be unable to keep pace with the the ionizing flux and the HII regions will therefore break out of the clouds \emph{in those directions}, even if they are confined in others. This process accounts in part for the very different structures of the wind bubbles as compared to their ionized counterparts. As well as being fundamentally better at blowing bubbles than the winds, there are almost always some vectors in each cloud in which this instability in the ionization front expansion is able to operate, so that the HII regions burst out along these vectors, ultimately exiting the clouds. This explains the more irregular appearance of the ionization--driven bubbles, and the fact that the HII regions seem to be able to penetrate further into the clouds than can the winds. The relative smoothness of the wind bubbles is then a consequence of the fact that, in clouds with an average density profile described by $\alpha=-2$, wind bubbles are stable (since their expansion is not accelerating) but HII regions are not.\\
\indent The corollary of this process is, however, that the HII regions almost inevitably rupture and begin venting their ionized gas outside the clouds, becoming, in effect, champagne flows. This in turn causes them to become flaccid, so that the main mechanism by which they are able to unbind the clouds -- accelerating cold gas beyond the system escape velocities by thermal pressure -- quickly begins to lose its effectiveness. It is highly likely that the same process will substantially decrease the effectiveness of pressure--driven winds as well.\\
\indent We show this qualitatively in Figure \ref{fig:HII_press} where we compare the measured mean pressure in the HII regions in Run I as a function of time with the expected fall--off in pressure in an HII region expanding in a uniform medium. The measured pressure is computed by averaging the pressure in the fifty percent of the ionized gas closest to the most massive ionized source. This is intended to exclude the very low density ionized gas which has escaped from the cloud. Note that we cannot compare to an HII region expanding in the correct density profile, since no such solution exists for the $\alpha=-2$ profile present here. This comparison is therefore somewhat artificial, but it does demonstrate that the pressure in the HII regions simulated here drops off abruptly and extremely quickly as the ionized gas bursts out from the confines of the cold gas. This in turn explains the observation that the rate at which the HII regions are able to unbind the clouds drops off soon after their expansion begins.\\
\indent The effect of winds on star formation is generally more negative than that of HII regions. The destructive effects of winds at small distances from the sources is high, which allows them to efficiently destroy the dense material in the main star--forming regions of the clouds and shut down star formation in these volumes. However, the ram pressure exerted by winds falls off rapidly with distance from the massive stars so that, as shown in Figure \ref{fig:pdf}, the winds are not able to generate large quantities of dense gas elsewhere in the clouds by sweeping up otherwise quiescent material. They are therefore not as good at triggering star formation as photoionization is, so the compensatory effects of induced star formation are very small in the case of winds acting alone, and the effect on the star formation rate and efficiency is overwhelmingly negative, if there is any effect at all.\\
\indent It is possible that these conclusions may be altered by the inclusion of pressure--driven, instead of momentum--driven winds. We crudely test this by taking advantage of the fact that the bubble expansion laws for momentum-- and pressure--driven winds are the same in clouds with $r^{-2}$ density profiles. We showed in Section 5.1 that the difference between the normalisation of the expansion laws between the two cases was roughly a factor of five. Since the normalizing factor in the momentum--driven case depends on $(\dot{M}v_{\infty})^{1/2}$, increasing the momentum flux by a factor of 25 approximates the affects of the fully adiabatic wind solution in the case of clouds with these density profiles. We therefore repeated Runs A, B and UB, on which momentum--driven winds have very little effect, including momentum fluxes enhanced by this factor. In Figure \ref{fig:ubclx25_plot} we plot the star formation efficiencies and unbound gas fractions as functions of time in the enhanced momentum flux UB calculation and the corresponding ionized calculation, compared to the fiducial wind calculation. Figure \ref{fig:ubclx25_snap} depicts column--density projections of the same three calculations after 1.2 Myr of evolution.\\
\indent The unbound mass fraction in the enhanced momentum flux case grows at a rate comparable to, and even somewhat faster than, the same quantity in the ionized calculation, and much faster than in the fiducial calculation. The star formation efficiency is less affected, but drops by a few percent relative to the fiducial simulation, also slightly lower than in the ionized run. The snapshots show that the reason for these changes is that the bubbles blown by the enhanced--momentum winds are somewhat larger and smoother than those generated by ionization in the same density field. The higher momentum--fluxes have also been more effective at destroying the filamentary gas in which most of the star formation is occurring. The results for the Runs A and B calculation with enhanced momentum flux were very similar.\\
\indent At first sight, this suggests that winds may be able to inflict damage on clouds comparable to that done by HII regions. However, we stress that the high momentum fluxes used in these calculations approximate fully adiabatic winds, which are unlikely to be realistic. The wind bubbles visible in the centre panel of Figure \ref{fig:ubclx25_snap} quickly burst out from the confines of the cold gas and, in reality, the $>10^{6}$K hot gas they would contain would very rapidly vent from the cloud, as the HII gas does in the corresponding ionized calculation (only much faster). This would in turn lead the wind bubbles to become severely under--pressured with respect to the simple approximation made here, again in common with the HII regions. Although the effect of winds on such clouds cannot be definitively settled until the dynamics of the hot gas can be modelled accurately, it is unlikely to be as great as that of ionization.\\
\section{Conclusions}
We have simulated the effects of O--star winds on a range of turbulent GMCs, making the simplifying assumption that the winds inject momentum only into the clouds. This simplification is reasonable in the geometries studied here for two reasons. Firstly, because very hot shocked wind gas would easily leak out of the highly non--uniform gas structures produced by turbulence, as does the ionized gas in our previous simulations, greatly reducing the influence of the wind pressure and facilitating cooling. Secondly, the expansion laws of momentum-- and pressure--driven wind bubbles in the $\rho\propto r^{-2}$ density profiles exhibited by our clouds are both linear in time, so that even if the wind bubbles were confined by the cold gas, which assumption is made about their driving agent should not greatly affect their behaviour.\\ 
\indent Our conclusions may be summarized as follows.\\
(i) In a given cloud and over a given time period of a few Myr, momentum--driven stellar winds acting alone have a considerably smaller influence than expanding HII regions on both the appearance and the dynamics of the cloud. In particular, winds unbind mass at a substantially slower rate than photoionization, since they are effective only in very dense gas at small size scales. Enhancing the wind momentum fluxes to simulate the action of adiabatic pressure--driven winds produces effects comparable to those of HII regions, but this is unlikely to be physical, since the winds should leak from the clouds in reality.\\
(ii) The bubbles produced by momentum--driven winds in realistic cloud environments are smaller and smoother than those produced by photoionization, largely owing to the instability of HII region expansion in steep density gradients, such as the $\rho\propto r^{-2}$ profiles which well--characterize the clouds studied here.\\
(iii) The effects of winds on the star formation process are almost always negative, in the sense of reducing the star formation efficiency, sometimes by factors approaching two. In clouds where winds are able to exert a significant dynamical influence, they are in general more effective in shutting down star formation than are HII regions because they are roughly as good at stopping accretion onto the main star--forming parts of the cloud, but are not as proficient at triggering star formation elsewhere.\\
(iv) In common with HII regions, the influence of winds is strongly affected by the escape velocity of the host clouds, with the higher--mass and higher--escape velocity clouds being largely immune to the effects of winds.\\
\section{Acknowledgements}
We thank the referee, Chris McKee, for very insightful comments and suggestions which enhanced the paper substantially. This research was supported by the DFG cluster of excellence `Origin and Structure of the Universe' (JED, BE). IAB acknowledges funding from the European Research Council for the FP7 ERC advanced grant project ECOGAL.\\

\bibliography{myrefs}

\label{lastpage}

\end{document}